\newif\ifOneCol

\ifOneCol
\documentclass[draftcls,onecolumn,11pt]{IEEEtran}
\else
\documentclass[twocolumn,10pt]{IEEEtran}
\fi

\usepackage[utf8]{inputenc}
\usepackage{graphicx}
\graphicspath{{graphics/}}
\usepackage{color}
\usepackage{array}
\usepackage{amssymb,amsmath}
\usepackage{mathrsfs} 

\allowdisplaybreaks 

\IEEEoverridecommandlockouts

\DeclareMathOperator{\erf}{erf}

\begin{document}

	\title{Distortion Distribution of Neural Spike Train\\ Sequence Matching with Optogenetics}
	
	\author{Adam Noel\IEEEauthorrefmark{1}, \IEEEmembership{Member, IEEE}, Dimitrios Makrakis, and Andrew W. Eckford, \IEEEmembership{Senior Member, IEEE}
		\thanks{Manuscript received August 31, 2017; revised February 22, 2018; accepted March 18, 2018. A preliminary version of this work was presented at IEEE GLOBECOM 2017 \cite{Noel2017d}. This work was supported in part by the Natural Sciences and Engineering Research Council of Canada (NSERC) through a postdoctoral fellowship. \emph{Asterisk indicates corresponding author}.}
		\thanks{\IEEEauthorrefmark{1}A.~Noel is with the School of Engineering, University of Warwick, Coventry, UK (email: adam.noel@warwick.ac.uk). He was previously with the School of Electrical Engineering and Computer Science, University of Ottawa, Ottawa, ON, Canada.}
		\thanks{D.~Makrakis is with the School of Electrical Engineering and Computer Science, University of Ottawa, Ottawa, ON, Canada.}
		\thanks{A.~W.~Eckford is with the School of Electrical Engineering and Computer Science, York University, Toronto, ON, Canada.}
		\thanks{\copyright~2017 IEEE. Personal use of this material is permitted. However, permission to use this material for any other purposes must be obtained from the IEEE by sending an email to pubs-permissions@ieee.org.}
		}

	\newcommand{\EXP}[1]{\exp\left(#1\right)}
	\newcommand{\ERF}[1]{\erf\left(#1\right)}
	
	\newcommand{\prob}[1]{P_\textnormal{#1}}
	
	\newcommand{\threshold}{\tau}
	
	\newcommand{\dt}{\Delta t}
	
	\makeatletter
	\newcommand{\vast}{\bBigg@{2.8}}
	\newcommand{\Vast}{\bBigg@{3.5}}
	\makeatother
	
	\newcommand{\light}{\ell}
	\newcommand{\rateTarget}{\lambda_\mathrm{T}}
	\newcommand{\probTarget}{g_\mathrm{T}}
	\newcommand{\fireThresh}{v_\textrm{fire}}
	\newcommand{\tMin}{t_\textrm{min}}
	\newcommand{\nMin}{{n_\textrm{min}}}
	\newcommand{\tSep}{t_i}
	\newcommand{\nSep}{n_\Delta}
	\newcommand{\distortion}{d}
	\newcommand{\distortionX}[1]{d_{#1}}
	
	\newcommand{\target}[1]{u_{#1}}
	\newcommand{\targetSeq}{\vec{u}}
	\newcommand{\targetLen}{M}
	\newcommand{\gen}[1]{v_{#1}}
	\newcommand{\genSeq}{\vec{v}}
	\newcommand{\genCand}[1]{w_{#1}}
	\newcommand{\genSeqCand}{\vec{w}}
	\newcommand{\genLen}{N}
	\newcommand{\offset}[1]{a_{#1}}
	\newcommand{\offsetSeq}{\vec{a}}
	\newcommand{\map}[1]{f\left(#1\right)}
	\newcommand{\mapDT}[1]{f\left[#1\right]}
	\newcommand{\kernel}[1]{h\left(#1\right)}
	\newcommand{\kernelDT}[1]{h\left[#1\right]}
	\newcommand{\kernelVal}[1]{h_{#1}}
	\newcommand{\kernelDTSq}[1]{h^2\left[#1\right]}
	\newcommand{\distortFcn}[1]{d\left(#1\right)}
	\newcommand{\distortFcnX}[2]{d_{#1}\left(#2\right)}
	\newcommand{\distortFcnXSq}[2]{d^2_{#1}\left(#2\right)}
	\newcommand{\distortFcnDT}[1]{d\left(#1\right)}
	\newcommand{\distortFcnSq}[1]{d^2\left(#1\right)}
	\newcommand{\distortFcnAvg}[1]{\overline{d}\left(#1\right)}
	\newcommand{\distortFcnXAvg}[2]{\overline{d_{#1}}\left(#2\right)}
	\newcommand{\distortFcnAvgSq}[1]{\overline{d}^2\left(#1\right)}
	\newcommand{\distortFcnXAvgSq}[2]{\overline{d_{#1}}^2\left(#2\right)}
	\newcommand{\distortVariance}[1]{\sigma^2_{#1}}
	\newcommand{\distortDeviation}[1]{\sigma_{#1}}
	\newcommand{\pulse}{\delta t}
	\newcommand{\pulseDT}{\Delta W}
	\newcommand{\pulseLen}{L}
	\newcommand{\pulseInd}{l}
	\newcommand{\dPulse}[1]{b_{#1}}
	\newcommand{\dPulseGen}[1]{c_{#1}}
	
	\newcommand{\bDelayRV}[1]{X_{#1}}
	\newcommand{\bDelayRVval}[1]{x_{#1}}
	\newcommand{\distortRV}{Y}
	\newcommand{\distortRVval}{y}
	\newcommand{\arbitraryRV}{Z}
	\newcommand{\arbitraryRVval}{z}
	\newcommand{\pmfFcn}[1]{p\left(#1\right)}
	
	\newcommand{\ion}{I_{\mathrm{on}}}
	
	
	\newcounter{mytempeqncnt}	
	
	\maketitle

	\begin{abstract}
		\emph{Objective:} This paper uses a simple optogenetic model to compare the timing distortion between a randomly-generated target spike sequence and an externally-stimulated neuron spike sequence. Optogenetics is an emerging field of neuroscience where neurons are genetically modified to express light-sensitive receptors that enable external control over when the neurons fire.
		\emph{Methods:} Two different measures are studied to measure the timing distortion. The first measure is the delay in externally-stimulated spikes. The second measure is the root mean square error between the filtered outputs of the target and stimulated spike sequences.
		\emph{Results:} The mean and the distribution of the distortion are derived in closed form when the target sequence generation rate is sufficiently low. The derived results are verified with simulations.
		\emph{Conclusion:} The proposed model and distortion measures can be used to measure the deviation between neuron spike sequences that are prescribed and what can be achieved via external stimulation.
		\emph{Significance:} Given the prominence of neuronal signaling within the brain and throughout the body, optogenetics has significant potential to improve the understanding of the nervous system and to develop treatments for neurological diseases.  This work is a step towards an analytical model to predict whether different spike trains were observed from the same external stimulus, and the broader goal of understanding the quantity and reliability of information that can be carried by neurons.
	\end{abstract}

	
	\section{Introduction}
	
	The nervous system is the most complex system of the human body, and understanding it is considered to be one of the biggest challenges in all of biology; see \cite[Ch.~45]{Sadava2014}. Its neural network creates up to $10^{14}$ connections within the brain and controls bodily functions such as muscle contraction. The transfer of information is not entirely internal; sensory neurons, such as those in the retina, generate and propagate signals in response to external stimuli.
	
	There is significant interest in developing methods to precisely control the external excitation of neurons, which could help improve our understanding of the nervous system and develop treatments for neurological diseases. One prominent example is the emerging field of optogenetics; see \cite{Fenno2011}. Optogenetics uses a relatively simple genetic modification to induce a neuron to express light-sensitive receptors on its membrane. These light-gated receptors can then be used to adjust the ion current across the membrane, which enables a light source to alter the membrane's electrical potential and control when it fires. Experiments in \cite{Nagel2002,Nagel2003} identified opsin-based receptors such as Channelrhodopsin (ChR) to be particularly suitable for optogenetic studies, due to its simplicity (requiring only one protein) and its compatibility for implanting in \emph{living} animals, as first demonstrated with the worm \emph{C. elegans} in \cite{Nagel2005}.
	
	From a communication perspective, nanoscale stimulators were proposed in \cite{Balasubramaniam2011,Mesiti2013} to control neurons and interface with a neural network. \cite{Wirdatmadja2016} proposed that such stimulators could be implemented using optogenetics and be implanted for long term use. More generally, the notion of precise neural control raises questions about the quantity and reliability of information that can be carried using neurons. Information-theoretic analysis of a single ChR receptor in \cite{Eckford2016a} showed that it has a remarkable capability to receive information. However, the information propagated by neurons is typically observed via the pulses that fire and not the behavior of individual receptors. Researchers typically measure the number and timing of fired pulses or ``spikes''; see \cite{Houghton2011,Lindner2016}. The importance of timing has been demonstrated in \cite{Srivastava2017}, where neural spike timing patterns in songbirds were manipulated with millisecond-scale variations to control respiratory behavior.
	
	As described in \cite{Houghton2011}, the timing of spikes does not need to be \emph{perfect} to carry information correctly; the same external stimulus can result in slightly different timing patterns in different neurons. Experimental data in \cite{Narayan2006} showed timing patterns with individual spikes that deviated by up to tens of milliseconds. Furthermore, spike timing patterns can change as they propagate along connected neurons; see \cite{Veletic2016a}. We are interested in the statistical modeling of deviations in spike sequences to assess how likely different sequences are carrying the same information. As a first step, in this paper we measure how effectively we could externally stimulate a spike train to match some desired or ``target'' spike train.
	
	In consideration of optogenetics, we model an ideal neuron that is charged by a light source under the integrate-and-fire model described in \cite{Abbott1999}. Although the integrate-and-fire neuron model is not as realistic as more robust models, such as those reviewed in \cite{Izhikevich2004}, it is a good starting point because its simplicity lends itself to analytical tractability.	We adapt the metrics-based approach for natural responses to external stimuli (reviewed in \cite{Houghton2011}) to compare the target and generated sequences by measuring the ``distance'' between them. In the context of this work, the distance between the sequences is a \emph{distortion} between the train that we can generate and the target train. In practice, for a certain distortion metric, a \emph{threshold distortion} should exist below which the pertinent information in the spike train can be recovered, i.e., a non-zero level of distortion should be acceptable. We only consider the generation of the spike train at the sensory neuron; we do not model the propagation of spikes along the neuron or their re-generation in adjacent neurons. Recent works in these directions include \cite{Ramezani} and \cite{Khodaei2016a,Veletic2016}, respectively.
	
	Since there are different ways to encode information in a spike train sequence, we are interested in studying different types of distortion metrics. However, to facilitate statistical analysis, we assume that the timing of the spikes in the target spike train is a memoryless process, i.e., a homogeneous Poisson process in continuous time. Spike trains have been modeled as or compared with homogeneous Poisson processes in \cite{Izhikevich2003,Wirdatmadja2016,Abbasi2015,Mesiti2013,Tezcan2015}. Spike trains have also been modeled as non-homogeneous Poisson processes (see \cite{Mesiti2013,Balevi2013,Amiri2017,Cacciapuoti2016,Veletic2016a}), such that the arrival rate is time-varying, but we focus on a homogeneous Poisson process for ease of analysis. From this perspective, we make the following contributions:
	\begin{enumerate}
		\item We study two classes of distortion metrics. First, we consider a simple delay-based metric, where we measure both the delay due to an individual spike and the total delay from the entire target spike train. Second, we consider filter-based metrics, which measure the timing distortion as the $\ell^\mathrm{p}$ norm between the filtered output of the target and generated sequences. We focus on the root mean square error (RMSE; i.e., the $\ell^2$ norm) and derive the distortion for \emph{any} target spike train.
		\item We derive the mean and distribution of the distortion, under the assumptions of memoryless target spike timing and a low target spike generation rate. We analytically derive the distribution of the delay of an individual spike and the RMSE with a filter that has 1 tap. We use normal approximations for the distributions of the cumulative delay and the RMSE with 2 or more filter taps.
		\item We verify our derivations by comparing our analytical results with simulated spike train sequences. Generally, our derivations are more accurate when the target spike generation rate is low relative to the charging time.
	\end{enumerate}
	
	Our work extends preliminary results that we presented in \cite{Noel2017d}. In \cite{Noel2017d}, we measured the RMSE distortion and its expected value for filters with 1 or 2 taps. We did not consider an arbitrary finite number of filter taps, delay distortion, or distortion distributions, as we do in this paper.
	
	The rest of the paper is organized as follows. Section~\ref{section_model} describes the neuron firing model. We analyze the delay distortion metric in Section~\ref{sec_delay}. We analyze the filter-based distortion metrics in Section~\ref{sec_filter}. We verify our derivations with simulations in Section~\ref{sec_results} and conclude this work in Section~\ref{sec_conclusions}.
	
	\section{System Model}
	\label{section_model}
	
	We consider a system consisting of a light source and a single neuron with multiple light-sensitive receptors on its surface. The light source can illuminate the neuron, which opens the receptor ion channels located on its surface and increases its internal potential until it fires. 
	
	For the sake of analysis, we will make several (mostly realistic) assumptions about this process:
	\begin{enumerate}
		\item The light source is either on or off, which can be appropriate for lasers or LEDs (both common in optogenetics; see \cite{Fenno2011}). We assume that the wavelength of light is within the excitation spectrum of the receptors\footnote{Future work could also consider the optical wavelengths used; opsins with different excitation spectra have been identified, and these could be used to create orthogonal signaling channels; see \cite{Klapoetke2014,Kawano2017}.}.
		\item The process of a receptor opening is stochastic (e.g., see \cite{Nagel2003}, or detailed analytical models in \cite{Nikolic2009}), but we assume that the overall current is equal to its expected plateau value $\ion$ when the light is on, which can be on the order of 100\,pA or more; see \cite{Nikolic2009,Grossman2011}. This assumption is appropriate if the number of receptors on the neuron is sufficiently large. 
		\item The neuron uses the integrate-and-fire model from \cite{Abbott1999} with capacitance $C$ and threshold $\tau$. Typical parameter values are $C=11.5\,$pF (see \cite{Grossman2011}) and $\tau=85\,$mV (see \cite{Izhikevich2004,Nelson2008}). The integrate-and-fire model provides analytical simplicity (i.e., we can assume a fixed charging time) at the expense of some fidelity. Alternative models, including those reviewed in \cite{Izhikevich2004}, capture the membrane potential dynamics in greater detail, e.g., modeling the refractory period after the neuron fires.
		\item Time is discretized into slots of $\dt$ that are shorter than the time necessary to charge the neuron. Specifically, there exists an integer $\nMin$ such that the integrate-and-fire threshold satisfies
		\begin{equation}
			\label{eqn:ThresholdCondition}
			\tau = \nMin \frac{I_{\mathrm{on}} \Delta t}{C}.
		\end{equation}
		
		In continuous time, the minimum firing time is $\tMin = \nMin \dt$, which is on the order of $10\,$ms or less when using the aforementioned parameter values. For analytical convenience, we apply the continuous time model when analyzing the delay-based distortion in Section~\ref{sec_delay} and the discrete time model when analyzing the filter-based distortion in Section~\ref{sec_filter}.
	\end{enumerate}

\begin{figure}[!tb]
		\centering
		\ifOneCol
			\includegraphics[width=0.5\linewidth]{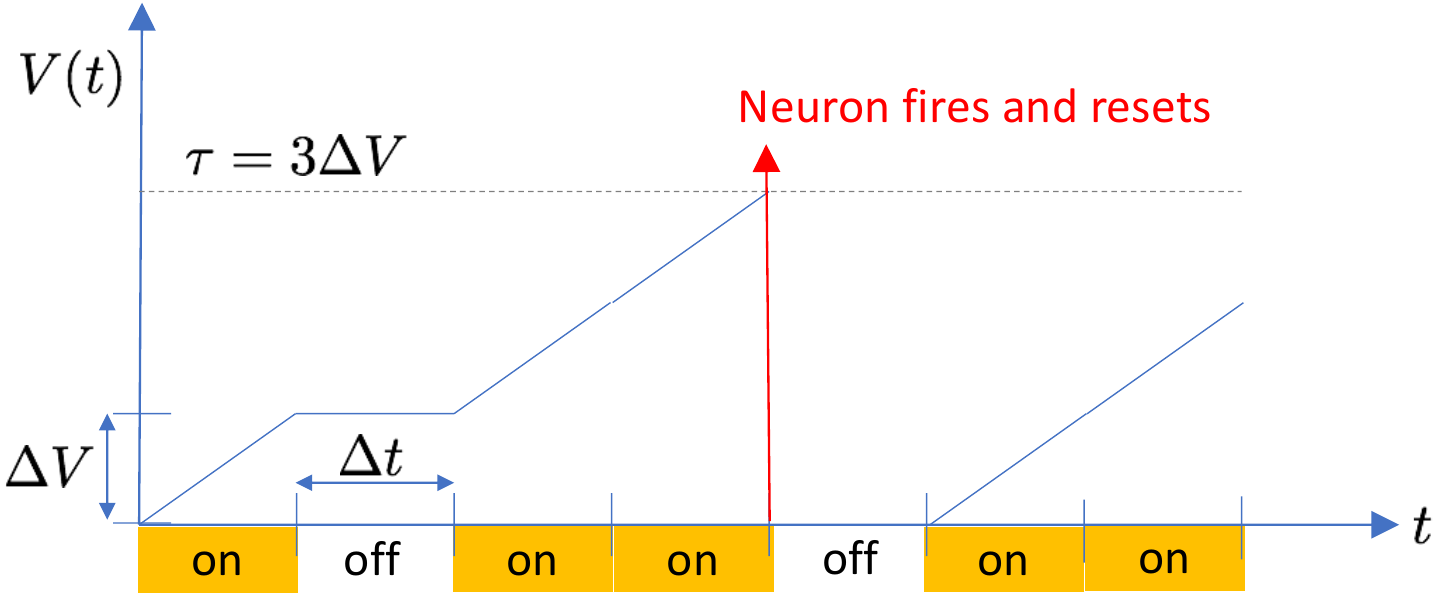}
		\else
			\includegraphics[width=\linewidth]{noel1}
		\fi
		\caption{Illustration of the neuron model with integrate-and-fire. In this example, $\nMin = 3$, i.e., spikes must be separated by at least $3 \Delta t$. In each interval when the light source is on, the voltage increases by $\Delta V$. Once the threshold $\tau = 3 \Delta V$ is reached, the neuron fires and resets to $V(t) = 0$. Even though this model can hold a charge indefinitely (see the second time step), this unrealistic feature is not critical in this paper because we stimulate the neuron continuously from its resting potential until it fires.}
		\label{fig:ModelDiagram}
	\end{figure}

	Based on these assumptions, our model is as follows:	
	\begin{itemize}		
	\item Let $V(t)$ represent the neuron potential (relative to its resting potential) as a function of time.
	From assumptions 1 to 3, when the light is on, the neuron behaves as an ideal capacitive circuit with a current $\ion$. Thus, if the light is on from time $t_1$ to $t_2$, then the change in potential $V(t_2) - V(t_1)$ is
	\begin{equation}
		\label{eqn:ChangeInPotentialContinuous}
		V(t_2)-V(t_1) = \frac{1}{C} \int_{t_1}^{t_2} I_{\mathrm{on}} \:dt = \frac{I_{\mathrm{on}} (t_2-t_1)}{C} .
	\end{equation}
	
	If the light is off, then the current is zero, so $V(t_2)-V(t_1) = 0$.
	
	\item From assumption 3, once $V(t)$ exceeds the threshold $\tau$, the neuron fires and $V(t)$ is immediately reset to zero\footnote{A neuron refractory period could be included by increasing $\nMin$ (or $\tMin$) as needed. The remaining analysis would be unaffected.}.
	
	\item From assumption 4,  we say in discrete-time that the light is synchronized with a clock and is either on or off for an entire interval $\Delta t$. Then from (\ref{eqn:ChangeInPotentialContinuous}) we define $\Delta V$ as
	\begin{equation}
		\label{eqn:DeltaV}
		\Delta V = V(t + \Delta t) - V(t) = \frac{I_{\mathrm{on}} \Delta t}{C},
	\end{equation}
	when the light source is on.
	Finally, since $\tau = \nMin \Delta V$ from (\ref{eqn:ThresholdCondition}), the light must be on for $\nMin$ slots in order for the neuron to fire. This is depicted in Fig. \ref{fig:ModelDiagram}. Analogously, in continuous time, the light must be on for $\tMin$ seconds for the neuron to fire.
	\end{itemize}
	
\begin{figure}[!tb]
		\centering
		\ifOneCol
		\includegraphics[width=0.5\linewidth]{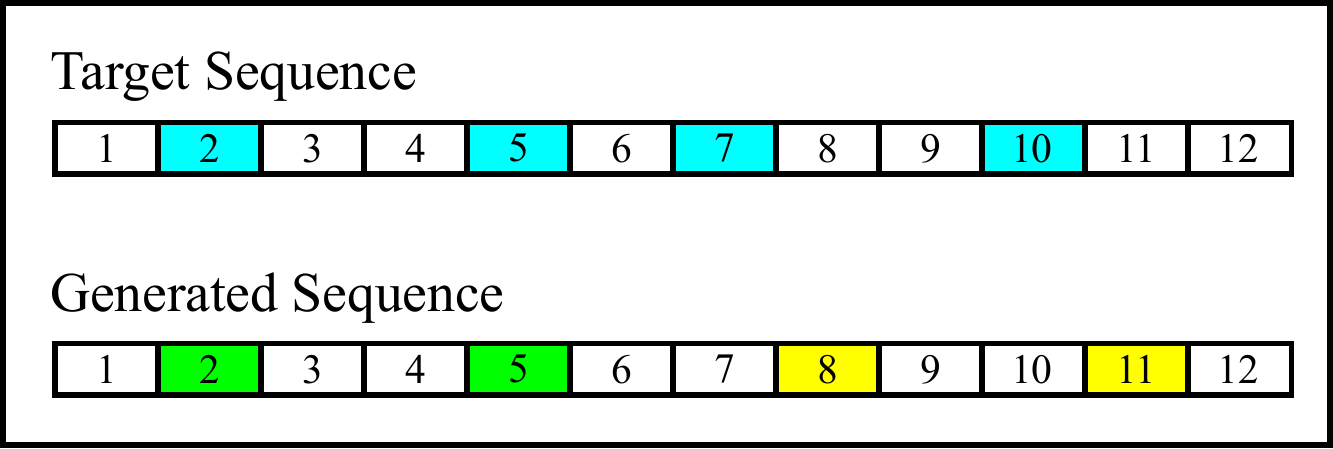}
		\else
		\includegraphics[width=\linewidth]{noel2}
		\fi
		\caption{Example of target sequence matching in discrete time, where the generated sequence is constrained by a charging time of $\nMin = 3$~slots. Slots are labeled chronologically and colored when there is a spike at the start of the slot. The target sequence has 4 pulses (shown in blue). The generated sequence can match the first 2 pulses (green, in the 2nd and 5th slots), but the final 2 pulses (yellow, in the 8th and 11th slots) have a delay of one slot.}
		\label{fig_matching_example}
	\end{figure}

	In this work, we are interested in using the light source to generate a train of spikes to match a target sequence, given that we are constrained by the time for the neuron to charge and fire. This matching problem is demonstrated in Fig.~\ref{fig_matching_example}. We assume there is a target spike train $\targetSeq$ that we define by the timing of its individual spikes, i.e., $\targetSeq = \{\target{1}, \target{2}, \ldots,\target{\targetLen}\}$, where $\target{i}$ is the time slot during which the $i$th spike fires. Without loss of generality, the firing times in $\targetSeq$ are in non-decreasing order. We assume that $\targetSeq$ is known \emph{a priori}. We may not be able to generate $\targetSeq$ perfectly, but instead we use the light source to generate the sequence $\genSeq = \{\gen{1}, \gen{2}, \ldots,\gen{\genLen}\}$. The only constraint on neuron firing times in $\genSeq$ that we consider is the time needed for the neuron to charge to the threshold voltage $\threshold$. Since $\targetSeq$ is known \emph{a priori}, we can turn on the light source and begin charging the neuron \emph{before} the corresponding target firing time. As long as a given target spike occurs at least $\nMin$ slots (or $\tMin$ seconds) after the previous spike, then we can generate a corresponding spike at the precise target time. Thus, if a sequence has a maximum spike generation frequency whose period is greater than $\tMin$, then we can generate the sequence \emph{perfectly}. This is a simplified and ideal generation model but it facilitates tractable analysis in this paper.
	
	If we are unable to generate the target sequence perfectly, then this does \emph{not} imply that $\targetSeq$ is physically unrealizable or that $\targetSeq$ does not contain biologically relevant information. The speed at which we can charge the neuron is a function of the overall current $I_{\mathrm{on}}$, which is constrained by the intensity of the light source and the number of light-sensitive receptors; see \cite{Nikolic2009}. Furthermore, it is common for neurons to multiplex signals from multiple input neurons (e.g., see \cite{Akam2014}), and in such a case it could be that the superposition of inputs leads to an ideal output signal that isn't realizable, even though the information is biologically relevant.
	
	Our goal in this paper is to measure the ``distance'' of the sequence $\genSeq$ from the sequence $\targetSeq$, subject to a distance or distortion measure $\distortFcn{\targetSeq,\genSeq}$. Obviously, if all spikes in $\targetSeq$ are separated by at least $\nMin$ slots or $\tMin$ seconds, then we can generate $\genSeq = \targetSeq$ and we should have $d(\targetSeq,\genSeq) = 0$. To make our formulation as general as possible, we do not always precisely describe how to generate $\genSeq$, but we usually assume that we can write the firing time of the $i$th spike in $\genSeq$ as
	\begin{equation}
	\gen{i} = \max\{\target{i}, \gen{i-1} + \nMin\}
	\label{eqn_generation}
	\end{equation}
	in discrete time, such that the sequences have the same length, and we replace $\nMin$ with $\tMin$ in continuous time. The initial spike in $\genSeq$ is always $\gen{1} = \target{1}$.
	
	An alternative approach to defining $\genSeq$ would be to consider a candidate sequence $\genSeqCand = \{\genCand{1}, \genCand{2}, \ldots,\genCand{\genLen}\}$. If $\mathcal{W}$ is the set of such sequences, then a necessary and sufficient condition for $\genSeqCand \in \mathcal{W}$ is that {\em spikes in $\genSeqCand$ are separated by at least $\tMin$ (in continuous time) or $\nMin$ (in discrete time)}. We could then optimize $\genSeq$ by finding the ``closest'' sequence in $\genSeqCand$, i.e.,
	\begin{equation}
		\genSeq = \arg \min_{\genSeqCand \in \mathcal{W}} \distortFcn{\targetSeq,\genSeqCand}.
	\end{equation}
	
	Such an optimization problem is distinct from (\ref{eqn_generation}) and might include skipping a spike in order to match other spikes (e.g., in Fig.~\ref{fig_matching_example}, we might not generate a pulse in the 8th slot in order to generate a pulse in the 10th slot). This would be an interesting problem for future work.

	\section{Delay Distortion Metric}
	\label{sec_delay}
	
	In this section, we assume (\ref{eqn_generation}) and measure the distortion as the delay between target spikes in $\targetSeq$ and the corresponding generated spikes in $\genSeq$. Thus, the distortion increases linearly with the delay. The distortion $\distortFcnX{i}{\target{i},\gen{i}}$ of the $i$th spike is
	\begin{equation}
	\distortFcnX{i}{\target{i},\gen{i}} = \gen{i} - \target{i},
	\label{eqn_delay}
	\end{equation}
	and, since $\gen{i} > \target{i}$, the distortion is always non-negative. We can immediately write the total (cumulative) delay distortion as $\distortFcn{\targetSeq,\genSeq} = \sum_{i=1}^\targetLen \distortFcnX{i}{\target{i},\gen{i}}$, where we assume that the lengths of the two sequences are equal, i.e., $\genLen = \targetLen$.
	
	In the remainder of this section, we derive the mean and distribution of the delay, where the spikes in $\genSeq$ follow (\ref{eqn_generation}). We first consider the delay associated with an individual spike in $\genSeq$. Then, we derive the total delay. We assume that the spikes in the target sequence $\targetSeq$ follow a Poisson process in continuous time with constant rate $\rateTarget$, but that the target generation rate is \emph{sufficiently low} such that the delay in generating the current pulse in $\genSeq$ is only a function of the generation time of the immediately previous pulse. Thus, there should be no more than two target pulses within any time interval of length $2\tMin$ (which is not strictly true for a real Poisson process) and the range of the delay for one spike is approximated as $[0,\tMin)$. The time separating consecutive target pulses is $\target{i}-\target{i-1} = \tSep$, which is an exponential random variable with mean $1/\rateTarget$.
	
	\subsection{Single Spike Delay}
	
	Based on our model, $\distortFcnX{1}{\target{1},\gen{1}} = 0$. The average delay $\distortFcnXAvg{i}{\target{i},\gen{i}}$ for any later spike is
	\begin{align}
	\distortFcnXAvg{i}{\target{i},\gen{i}} = &\,\mathbf{E}\{\distortFcnX{i}{\target{i},\gen{i}}\} \nonumber \\
	= & \Pr\{\tSep < \tMin\} \cdot \mathbf{E}\{\tMin-\tSep | \tSep < \tMin\} \nonumber \\
	\approx &\, (1-\exp(-\rateTarget\tMin)) \nonumber \\
	&\times[\tMin - \mathbf{E}\{\tSep | \tSep < \tMin\}],
	\label{eqn_delay_expected}
	\end{align}
	where $\mathbf{E}\{\cdot\}$ is the expectation, (\ref{eqn_delay_expected}) accounts for the fact that $\distortionX{i} = 0$ if $\tSep \ge \tMin$, and the approximation in (\ref{eqn_delay_expected}) and the remaining equations in this section comes from the assumption that $\rateTarget$ is small. From the properties of conditional expectation (see \cite[Ch.~2]{Proakis2000}) and integration by parts, we can write
	\begin{align}
	\mathbf{E}\{\tSep | \tSep < \tMin\} = &\, \frac{p_{\tSep,\tSep < \tMin}(\tSep,\tSep < \tMin)}{\Pr\{\tSep < \tMin\}} \nonumber \\
	\approx &\, \frac{\int_0^{\tMin}\tSep\rateTarget\exp(-\rateTarget\tSep)\mathrm{d}\tSep}{1-\exp(-\rateTarget\tMin)} \nonumber \\
	= &\, \frac{\frac{1}{\rateTarget} - \left(\tMin + \frac{1}{\rateTarget}\right)\exp(-\rateTarget\tMin)}{1-\exp(-\rateTarget\tMin)},
	\label{eqn_delay_expected_given_nonzero}
	\end{align}
	where $p_{\tSep,\tSep < \tMin}(\tSep,\tSep < \tMin)$ is the joint probability that the time between successive pulses has value $\tSep$ \emph{and} that $\tSep < \tMin$. We can substitute (\ref{eqn_delay_expected_given_nonzero}) into (\ref{eqn_delay_expected}) and write the average delay as
	\begin{equation}
	\label{eqn_delay_expected_evaluated}
	\distortFcnXAvg{i}{\target{i},\gen{i}}
	\approx \tMin + \frac{1}{\rateTarget}\left(\exp(-\rateTarget\tMin)-1\right),
	\end{equation}
	which again we emphasize applies when $i > 1$.
	
	We can use the same assumptions to write the distribution of $\distortFcnX{i}{\target{i},\gen{i}}$. From the properties of conditional expectation and exponential random variables, it can be shown that the cumulative distribution function (CDF) of the distortion is
	\begin{equation}
	\label{eqn_delay_distribution_single}
	\Pr\{\distortFcnX{i}{\target{i},\gen{i}} \le \distortRVval\} \approx \left\{
	\begin{array}{cl}
	0, & \distortRVval < 0 \\
	\EXP{-\rateTarget(\tMin-\distortRVval)}, & \distortRVval \in [0,\tMin]\\
	1, & \distortRVval > \tMin.
	\end{array}
	\right. 
	\end{equation}
	
	\subsection{Total Sequence Delay}
	
	From (\ref{eqn_delay_expected_evaluated}), we write the \emph{total} expected delay $\distortFcnAvg{\targetSeq,\genSeq}$ as
	\begin{equation}
	\label{eqn_delay_expected_total}
	\distortFcnAvg{\targetSeq,\genSeq} \approx (\targetLen-1)\left[\tMin + \frac{1}{\rateTarget}\left(\exp(-\rateTarget\tMin)-1\right)\right].
	\end{equation}
	
	The distribution of the total delay is not as readily tractable as the distribution of a single spike. If we assume that all of the delays are independent, the sum of the delays is still a combinatorial problem due to the piecemeal behavior of each delay's distribution (see (\ref{eqn_delay_distribution_single})). Instead, it is more convenient to consider a normal approximation, where we assume independent delays \emph{and} that the number of spikes is sufficiently large to apply the central limit theorem. First, we need the second moment of a single spike's delay, i.e.,
	\begin{align}
	\mathbf{E}\{\distortFcnXSq{i}{\target{i},\gen{i}}\} = & \Pr\{\tSep < \tMin\} \cdot \mathbf{E}\{(\tMin-\tSep)^2 | \tSep < \tMin\} \nonumber \\
	\approx & \, (1-\EXP{-\rateTarget\tMin}) \nonumber \\
	&\times\big[\tMin^2 + \mathbf{E}\{\tSep^2 | \tSep < \tMin\} \nonumber \\
	&- 2\tMin\mathbf{E}\{\tSep | \tSep < \tMin\}\big],
	\label{eqn_delay_second_moment}
	\end{align}
	where the first moment of $\tSep$ is given by (\ref{eqn_delay_expected_given_nonzero}), and we can similarly derive the second moment of $\tSep$ as
	\begin{align}
	\mathbf{E}\{\tSep^2 | \tSep < \tMin\} \approx &\, \frac{\int_0^{\tMin}\tSep^2\rateTarget\exp(-\rateTarget\tSep)\mathrm{d}\tSep}{1-\exp(-\rateTarget\tMin)} \nonumber \\
	= &\, \frac{\frac{2}{\rateTarget} - \left(\tMin^2 + \frac{2}{\rateTarget} + \frac{2}{\rateTarget^2}\right)\EXP{-\rateTarget\tMin}}{1-\EXP{-\rateTarget\tMin}}.
	\label{eqn_tSep_second_moment}
	\end{align}
	
	By substituting (\ref{eqn_delay_expected_given_nonzero}) and (\ref{eqn_tSep_second_moment}) into (\ref{eqn_delay_second_moment}), the second moment of $\distortFcnXSq{i}{\target{i},\gen{i}}$ becomes
	\begin{equation}
	\mathbf{E}\{\distortFcnXSq{i}{\target{i},\gen{i}}\} \approx \tMin^2 + \frac{2}{\rateTarget}\left(\frac{1}{\rateTarget} - \tMin\right) - \frac{2\EXP{-\rateTarget\tMin}}{\rateTarget^2},
	\end{equation}
	thus, the variance of $\distortFcnX{i}{\target{i},\gen{i}}$, $\distortVariance{\distortionX{i}}$, is
	\begin{align}
	\distortVariance{\distortionX{i}} = &\, \mathbf{E}[\distortFcnXSq{i}{\targetSeq,\genSeq}] - \distortFcnXAvgSq{i}{\targetSeq,\genSeq} \nonumber \\
	\approx &\, \frac{1}{\rateTarget}\left[1 - \EXP{-2\rateTarget\tMin}\right] - \frac{2\tMin\EXP{-\rateTarget\tMin}}{\rateTarget},
	\end{align}
	and the variance of the total delay is $\distortVariance{\distortion} = (\targetLen-1)\distortVariance{\distortionX{i}}$. We can then use the normal approximation to write the CDF of the total delay as
	\begin{equation}
	\label{eqn_distortion_cdf_total}
	\Pr(\distortFcnDT{\targetSeq,\genSeq} \le \distortRVval)
	\approx \frac{1}{2}\left[1 + \ERF{\frac{\distortRVval - \distortFcnAvg{\targetSeq,\genSeq}}{\sqrt{2\distortVariance{\distortion}}}}\right].
	\end{equation}
	
	The accuracy of (\ref{eqn_distortion_cdf_total}) is sensitive to both the accuracy of the average delay (which assumes a low generation rate $\rateTarget$) and the length of the sequence $\targetLen$.
	
	\section{Filter-Based Distortion Metrics}
	\label{sec_filter}
	
	In this section, we present the filter-based metric model from \cite{Houghton2011} for comparing spike trains and apply it to measure the $\ell^\mathrm{p}$ norm of the timing distortion between the target sequence $\targetSeq$ and the generated sequence $\genSeq$. For tractability we use the discrete time model. We focus on the RMSE of the timing and derive both the mean and the distribution of this distortion for finite filter lengths.
	
	\subsection{Filter-Based Model from \cite{Houghton2011}}
	
	The delay distortion is a rather simple metric to distinguish between pairs of spike sequences. For example, the delay distortion increases linearly with the length of the delay. To enable more discretion in how to measure the distortion, we consider \emph{filter-based metrics}, as presented in \cite{Houghton2011}. These metrics map spike trains onto the vector space of functions via a filter kernel, and we can choose the kernel. The length and structure of the kernel controls the sensitivity of the distortion metric. \cite{Houghton2011} also presented \emph{edit distance} metrics, which use operations to convert one spike sequence into the other. Additional alternatives include \emph{pairwise correlation}, \emph{Hunter-Milton similarity}, \emph{event synchronization}, and \emph{stochastic event synchrony}, as summarized in \cite{Dauwels2008}. We focus here on the filter-based metrics due to their performance with experimental data (as shown in \cite{Houghton2011,Narayan2006}) and their analytical tractability.	
	
	We begin by mapping the spike trains $\targetSeq$ and $\genSeq$ onto the vector space of functions. A discrete time model of the $\ell^\mathrm{p}$ norm distortion in \cite[Eq.~(17)]{Houghton2011} between $\targetSeq$ and $\genSeq$ is
	\begin{equation}
		\label{eqn_distortionDT_lp_norm}
		\distortFcnDT{\targetSeq,\genSeq} = \left(\sum_n\left|\mapDT{n;\targetSeq} - \mapDT{n;\genSeq}\right|^\mathrm{p} \right)^{1/\mathrm{p}},
	\end{equation}
	where $n$ is the time index and $\mapDT{n;\cdot}$ is the mapping function that maps a sequence to a vector space.  We use a filter function with a kernel $\kernelDT{n}$ and length $\pulseDT$, such that the sequence $\targetSeq$ maps as
	\begin{equation}
		\label{eqn_filter_function}
		\mapDT{n;\targetSeq} = \sum_{i=1}^\targetLen \kernelDT{n-\target{i}}.
	\end{equation}
	
	For ease of analysis, we are interested in kernels with a finite number of taps. Other kernels considered in \cite{Houghton2011} include the Gaussian filter and the exponential filter, but they are outside the scope of this work. The exponential filter was also used in \cite{Narayan2006}, where it was shown that stimuli could be classified with reasonable accuracy even though the average RMSE between sequences from the same stimulus with dozens of spikes could be as high as 25.
		
	\subsection{Filter-Based Metric with RMSE}
	
	We focus on analyzing the $\ell^2$ norm, i.e., the Euclidean distance or root mean square error (RMSE) between the two sequences in vector space, which we will find is sensitive to the timing of the individual spikes in $\targetSeq$ and $\genSeq$. From (\ref{eqn_distortionDT_lp_norm}) and (\ref{eqn_filter_function}), the distortion can be written as
	\begin{align}
		\distortFcnDT{\targetSeq,\genSeq} = & \left(\sum_n\left(\sum_{i=1}^\targetLen \kernelDT{n-\target{i}} - \sum_{i=1}^\genLen \kernelDT{n-\gen{i}}\right)^2\right)^\frac{1}{2} \nonumber \\
		= & \vast(\sum_n\vast[\left(\sum_{i=1}^\targetLen \kernelDT{n-\target{i}}\right)^2 + \left(\sum_{i=1}^\genLen \kernelDT{n-\gen{i}}\right)^2 \nonumber \\
		& - 2\sum_{i=1}^\targetLen \sum_{j=1}^\genLen \kernelDT{n-\target{i}}\kernelDT{n-\gen{j}} \vast]\vast)^\frac{1}{2} \nonumber \\
		= & \vast(\underbrace{\sum_{i=1}^\targetLen \sum_n\kernelDTSq{n-\target{i}}}_\text{Target energy} + \underbrace{\sum_{i=1}^\genLen\sum_n \kernelDTSq{n-\gen{i}}}_\text{Generated energy} \nonumber \\
		& + \underbrace{2\sum_{i=1}^\targetLen\sum_{j=i+1}^\targetLen\sum_n \kernelDT{n-\target{i}} \kernelDT{n-\target{j}}}_\text{Target density} +  \nonumber \\
		& + \underbrace{2\sum_{i=1}^\genLen\sum_{j=i+1}^\genLen\sum_n \kernelDT{n-\gen{i}} \kernelDT{n-\gen{j}}}_\text{Generated density} \nonumber \\
		& - \underbrace{2\sum_{i=1}^\targetLen \sum_{j=1}^\genLen\sum_n \kernelDT{n-\target{i}}\kernelDT{n-\gen{j}}}_\text{Overlap measure} \vast)^\frac{1}{2}.
		\label{eqn_l2_distortion}
	\end{align}

	In (\ref{eqn_l2_distortion}), we label each of the terms that comprise $\distortFcnDT{\targetSeq,\genSeq}$. The two \emph{energy} terms describe the energy of the two filtered sequences. The \emph{density} terms describe the proximity of the individual spikes in each sequence to the other spikes in the \emph{same} sequence. The \emph{overlap measure} describes the proximity of the individual spikes in $\genSeq$ to the spikes in $\targetSeq$. It can be shown, as expected, that the distortion is minimized to $d(\targetSeq,\genSeq) = 0$ when the overlap measure is maximized, i.e., when $\genSeq = \targetSeq$.
	
	We have not yet placed any constraints on the forms of $\targetSeq$, $\genSeq$, or the kernel $\kernelDT{n}$. For simplification, we now impose that we generate a sequence of the same length as the target sequence, i.e., $\genLen = \targetLen$, such that we can write the timing of each spike in $\genSeq$ as $\gen{i} = \target{i} + \offset{i}$, where $\offset{i}$ is the offset of the $i$th generated spike from the target time. This is still more general than the generation model in (\ref{eqn_generation}), but is sufficient for us to combine the sequence energy terms and write the distortion as
	\begin{align}
		\distortFcnDT{\targetSeq,\genSeq} = 
		& \vast(2\sum_{i=1}^\targetLen \sum_n\kernelDTSq{n-\target{i}} \nonumber \\
		& +  2\sum_{i=1}^\targetLen\sum_{j=i+1}^\targetLen\Bigg(\sum_n \kernelDT{n-\target{i}} \kernelDT{n-\target{j}} \nonumber \\
		& + \sum_n \kernelDT{n-\target{i}-\offset{i}} \kernelDT{n-\target{j}-\offset{j}}\Bigg) \nonumber \\
		& - 2\sum_{i=1}^\targetLen \sum_{j=1}^\targetLen\sum_n \kernelDT{n-\target{i}}\kernelDT{n-\target{j}-\offset{j}} \vast)^\frac{1}{2}.
		\label{eqn_l2_distortion_offset}
	\end{align}
	
	Our notion of ``proximity'' between spikes when measuring the overlap or density of sequences is particularly sensitive to the length of the kernel $\kernelDT{n}$, i.e., the number of filter taps. To explore this further, we consider a kernel of length $\pulseDT = 1$ discrete time slot before generalizing to any finite length filter. Furthermore, we assume that the spike times in the target sequence $\targetSeq$ are all unique (i.e., there is no more than one target spike in a given slot), which is practical given that we will also assume a low target spike generation rate.
	
	\subsection{RMSE with Kernel of Length 1}
	\label{sec_distortion_kernel1}
	
	If $\pulseDT = 1$, then from (\ref{eqn_l2_distortion}) the target density term must be 0, i.e., there is no partial overlap between filtered spikes in the same sequence. Every spike in $\genSeq$ either perfectly matches or misses a target spike; partial overlap between the two sequences is not possible. Relative to other kernel lengths, the \emph{overlap} term is minimized. From the perspective of matching $\targetSeq$ with $\genSeq$, this distortion measure discards every generated spike that does not align perfectly with a target spike.
	
	By applying a Kronecker delta kernel, where $\kernelDT{n} = \delta[n]$, the distortion in (\ref{eqn_l2_distortion_offset}) becomes
	\begin{equation}
		\distortFcnDT{\targetSeq,\genSeq} = \bigg(2\targetLen - 2\sum_{i=1}^\targetLen \sum_{j=1}^\targetLen \left(\target{i} \stackrel{?}{=} \target{j} + \offset{j}\right) \bigg)^\frac{1}{2},
		\label{eqn_l2_distortion_simplified}
	\end{equation}
	where $\big(\target{i} \stackrel{?}{=} \target{j} + \offset{j}\big)$ is an indicator function with value 1 when it is true. Eq.~(\ref{eqn_l2_distortion_simplified}) has a form that we can readily evaluate when given a specific target sequence $\targetSeq$ and an offset sequence $\offsetSeq = \{\offset{1},\offset{2},\ldots,\offset{\targetLen}\}$. We emphasize that (\ref{eqn_l2_distortion_simplified}) makes no assumptions about the offset value $\offset{j}$ (except that each $\offset{j}$ term is an integer), and it accounts for generated spikes that align with \emph{any} target spike (not only the corresponding target). However, the neuron must be re-charged after every spike generation, thus the indicator function can only be true for at most one value of $j$ for every value of $i$ (and vice versa).
	
	We now consider the distribution of the distortion measure $\distortFcnDT{\targetSeq,\genSeq}$. For tractability, we make additional assumptions about the target sequence and the offset sequence. We impose that each offset $\offset{j}$ must be a \emph{delay}, i.e., $\offset{j} \ge 0$, and the firing times in $\genSeq$ follow (\ref{eqn_generation}). We also assume that the target sequence is sparse, such that it has no more than 2 spikes within any interval of $2\nMin$ slots, where $\nMin$ is the minimum number of slots between spikes in $\genSeq$. Thus, each $\offset{j}$ will only depend on the values of the corresponding $\target{j}$ and $\target{j-1}$, such that we are never waiting to generate more than 1 spike at a time. By imposing these assumptions, a generated spike that occurs at the same time as a target spike must be intended for that target, and we can approximate the distortion in (\ref{eqn_l2_distortion_simplified}) as
	\begin{equation}
		\distortFcnDT{\targetSeq,\genSeq} \approx \bigg(2\targetLen - 2\sum_{i=1}^\targetLen \left(\offset{i} \stackrel{?}{=} 0\right) \bigg)^\frac{1}{2},
		\label{eqn_l2_distortion_simplified_low}
	\end{equation}
	where the approximation is due to the sparsity assumption.
	
	Let us consider whether the assumptions for (\ref{eqn_l2_distortion_simplified_low}) prevent us from satisfying $\big(\target{i} \stackrel{?}{=} \target{j} + \offset{j}\big)$ in (\ref{eqn_l2_distortion_simplified}) when $\offset{i} \ne 0$. We can prove by contradiction that this is true. If $\offset{i} \ne 0$, then $\big(\target{i} \stackrel{?}{=} \target{j} + \offset{j}\big)$ could only be true for some $i \ne j$. We've imposed that $\offset{i}$ must be non-negative, so we can only consider $j < i$. The most recent case that could satisfy the indicator function would be $j = i - 1$, such that $\target{i} = \target{i-1} + \offset{i-1}$. However, if $\offset{i-1} \ne 0$, then $\target{i-1} - \target{i-2} < \nMin$, and the timing of the $(i-2)$th, $(i-1)$th, and $i$th spikes violates our assumption that there can be no more than 2 target spikes within any interval of $2\nMin$ slots. Consequently, we only need to look for cases of $\offset{i} = 0$, which leads to (\ref{eqn_l2_distortion_simplified_low}).
	
	From the approximate distortion in (\ref{eqn_l2_distortion_simplified_low}), we can find the distribution and the expected value. First, we need the probability that $\offset{i} = 0$. Since $\targetSeq$ is increasing, the first offset $\offset{1} = 0$. For $i>1$, we know that $\offset{i} = 0$ if there is sufficient separation between the current and previous target spikes, i.e., if $\target{i} - \target{i-1} \ge \nMin$. In analogy with a Poisson process in continuous time, we assume that the number of slots separating consecutive target spikes follows a geometric distribution with probability $\probTarget$; see \cite{Ross2009}. We also assume that $\probTarget$ is small, since we assumed earlier that the target sequence is sparse. Thus, we can estimate the probability that $\offset{i} = 0, i > 1$, as
	\begin{align}
		\Pr(\offset{i} = 0) \approx & \Pr(\target{i} - \target{i-1} \ge \nMin) \nonumber \\
		= & \left(1-\probTarget\right)^{\nMin-1},
		\label{eqn_l2_distortion_cdf}
	\end{align}
	where the approximation is due to the assumption that $\probTarget \ll 1$, such that the probability that $\offset{i} = 0$ is a Bernoulli random variable with success probability $\left(1-\probTarget\right)^{\nMin-1}$. Furthermore, the summation $\bDelayRV{}= \sum_{i=2}^\targetLen(\offset{i} \stackrel{?}{=} 0)$, which is the number of target spikes (after the initial spike) we can generate with no delay, is a Binomial random variable with $\targetLen-1$ trials and value $\bDelayRVval{}$. Using the properties of functions of random variables (see \cite[Ch.~2]{Proakis2000}), we can then write the \emph{expected} distortion as
	\begin{align}
		\distortFcnAvg{\targetSeq,\genSeq} = &\, \mathbf{E}\left[\distortFcnDT{\targetSeq,\genSeq}\right] \approx
		\sum_{\bDelayRVval{}=0}^{\targetLen-1} \left(2\targetLen - 2 - 2\bDelayRVval{} \right)^\frac{1}{2} \pmfFcn{\bDelayRVval{}} \nonumber \\
		= & \sum_{\bDelayRVval{}=0}^{\targetLen-1} \left(2\targetLen - 2 - 2\bDelayRVval{} \right)^\frac{1}{2}\binom{\targetLen-1}{\bDelayRVval{}}\left(1-\probTarget\right)^{\bDelayRVval{}(\nMin-1)} \nonumber \\
		&\times  \left(1-\left(1-\probTarget\right)^{\nMin-1}\right)^{\targetLen-1-\bDelayRVval{}},
		\label{eqn_l2_distortion_mean}
	\end{align}
	where $\pmfFcn{\bDelayRVval{}}$ is the probability mass function (PMF) of the random variable $\bDelayRV{}$. If we write the distortion $\distortFcnDT{\targetSeq,\genSeq}$ as a random variable with value $\distortRVval$, then the cumulative density function (CDF) of the distortion is
	\begin{align}
		\Pr(\distortFcnDT{\targetSeq,\genSeq} \le \distortRVval)
		\approx & \Pr\left(\left(2\targetLen - 2 - 2\bDelayRVval{} \right)^\frac{1}{2} \le \distortRVval\right) \nonumber \\
		= & \Pr\left(\bDelayRVval{} \ge \targetLen - 1 - \frac{\distortRVval^2}{2}\right) \nonumber \\
		= &\, I_{(1-\probTarget)^{\nMin-1}}\left(\targetLen - 1 - \frac{\distortRVval^2}{2}, 1 + \frac{\distortRVval^2}{2}\right),
		\label{eqn_l2_distortion_distribution}
	\end{align}
	where $I$ is the regularized incomplete beta function; see \cite[Eq.~(6.6.2)]{Abramowitz1964}.
	
	\subsection{RMSE with Kernel of Arbitrary Finite Length}
	\label{sec_distortion_kernelX}
	
	If $\pulseDT > 1$, then partial overlap between spikes is possible, and the degree of distortion is less sensitive to the precise alignment of the spikes. For example, let us consider the arbitrary case $\pulseDT = \pulseLen$, such that the kernel is
	\begin{equation}
		\kernelDT{n} = \sum_{\pulseInd=0}^{\pulseLen-1} \kernelVal{\pulseInd}\delta[n-\pulseInd].
	\end{equation}
	
	A degree of overlap within a sequence can now occur between any pair of spikes, so we cannot simplify the exact expression for distortion in (\ref{eqn_l2_distortion_offset}) without approximations. We assume that the target sequence is sufficiently sparse so that overlap can only occur between consecutive spikes, i.e., the $i$th spike can overlap the $(i+1)$th spike but not the $(i+2)$th. In other words, we assume that there are no more than 2 target spikes within any interval of $2(\nMin + \pulseLen-1)$ slots. Then, we can write the density of the $i$th and $(i+1)$th spikes in $\targetSeq$ as
	\begin{equation}
		\sum_n \kernelDT{n-\target{i}} \kernelDT{n-\target{i+1}} =
		\sum_{n = \dPulse{i}}^{\pulseLen-1} \kernelVal{n}\kernelVal{n-\dPulse{i}},
		\label{eqn_target_density}
	\end{equation}
	if and only if the $i$th and $(i+1)$th spikes are separated by $\dPulse{i} \in [1,\pulseLen-1]$ slots. An analogous expression can be written for the density of consecutive spikes in the generated sequence $\genSeq$, though it only applies if the kernel length is longer than the minimum charging time, i.e., if $\pulseLen > \nMin$. Furthermore, from the sparsity assumption and by assuming the generation model in (\ref{eqn_generation}), the only slots that can have overlap between the filtered sequences $\targetSeq$ and $\genSeq$ is when a spike can be generated at the same time as its corresponding target with no delay, i.e., when $\offset{i} = 0$. Finally, we also repeat our assumption that the number of slots separating consecutive target spikes follows a geometric distribution with probability $\probTarget$. Applying all of these assumptions to (\ref{eqn_l2_distortion_offset}) leads to
	\begin{align}
		\distortFcnDT{\targetSeq,\genSeq} \approx 
		& \vast(2\targetLen \sum_{\pulseInd=0}^{\pulseLen-1}\kernelVal{\pulseInd}^2 +  2\sum_{i=1}^{\targetLen-1}\Bigg[(\dPulse{i} \stackrel{?}{<} \pulseLen) \sum_{n = \dPulse{i}}^{\pulseLen-1} \kernelVal{n}\kernelVal{n-\dPulse{i}} \nonumber \\
		& + (\dPulseGen{i} \stackrel{?}{<} \pulseLen) \sum_{n = \dPulseGen{i}}^{\pulseLen-1} \kernelVal{n}\kernelVal{n-\dPulseGen{i}} \nonumber\Bigg] \nonumber \\
		& - 2\sum_{\pulseInd=0}^{\pulseLen-1}\kernelVal{\pulseInd}^2 \sum_{i=1}^{\targetLen}(\offset{i} \stackrel{?}{=} 0) \vast)^\frac{1}{2},
		\label{eqn_l2_distortion_offset_simplified_widthX}
	\end{align}
	where the $i$th and $(i+1)$th spikes in the generated sequence $\genSeq$ are separated by $\dPulseGen{i}$. For tractability in the statistical analysis, we will find it useful to assume that $\dPulseGen{i} \ge \pulseLen, \forall i$, i.e., we ignore the density of the generated sequence (which is valid anyway if $\pulseLen \le \nMin$). This assumption enables us to simplify (\ref{eqn_l2_distortion_offset_simplified_widthX}) as 
	\begin{align}
		\distortFcnDT{\targetSeq,\genSeq} \approx 
		& \vast(2\targetLen \sum_{\pulseInd=0}^{\pulseLen-1}\kernelVal{\pulseInd}^2 +  2\sum_{i=1}^{\targetLen-1}(\dPulse{i} \stackrel{?}{<} \pulseLen) \sum_{n = \dPulse{i}}^{\pulseLen-1} \kernelVal{n}\kernelVal{n-\dPulse{i}} \nonumber \\
		& - 2\sum_{\pulseInd=0}^{\pulseLen-1}\kernelVal{\pulseInd}^2 \left[1 + \sum_{i=2}^{\targetLen}(\target{i} - \target{i-1} \stackrel{?}{\ge} \nMin)\right] \vast)^\frac{1}{2}.
		\label{eqn_l2_distortion_offset_simplified2_widthX}
	\end{align}
	
	Unlike the case where $\pulseDT=1$, we cannot readily write the distribution of this distortion in an analytically tractable form. However, we can determine the expected value and use a normal approximation for the distribution. To write the expected value of (\ref{eqn_l2_distortion_offset_simplified2_widthX}), we need to consider $\pulseLen$ random variables. There are $\pulseLen-1$ random variables of the form $\bDelayRV{\dPulse{}}$, which is the number of target spikes that are $\dPulse{}$ slots after the previous spike, i.e.,
	\begin{equation}
		\bDelayRV{\dPulse{}} = \sum_{i=1}^{\targetLen-1}(\target{i+1} - \target{i} \stackrel{?}{=} \dPulse{}),
	\end{equation}
	and each random variable of this form is a Binomial random variable with $\targetLen-1$ trials, success probability $(1-\probTarget)^{\dPulse{}-1}\probTarget$, and value $\bDelayRVval{\dPulse{}}$. The other random variable, $\arbitraryRV$, is the number of target spikes that are separated by at least $\nMin$ slots, i.e.
	\begin{equation}
		\arbitraryRV = \sum_{i=2}^{\targetLen}(\target{i} - \target{i-1} \stackrel{?}{\ge} \nMin),
	\end{equation}
	which is a Binomial random variable with $\targetLen-1$ trials, success probability $(1-\probTarget)^{\nMin-1}$, and value $\arbitraryRVval$. Using the random variables, we can rewrite (\ref{eqn_l2_distortion_offset_simplified2_widthX}) as
	\begin{equation}
		\distortFcnDT{\targetSeq,\genSeq} \approx \!\! \left[2(\targetLen-1-\arbitraryRV)\sum_{\pulseInd=0}^{\pulseLen-1}\kernelVal{\pulseInd}^2 + 2\sum_{\dPulse{}=1}^{\pulseLen-1}\bDelayRV{\dPulse{}}\sum_{n = \dPulse{}}^{\pulseLen-1} \kernelVal{n}\kernelVal{n-\dPulse{}}\right]^\frac{1}{2}.
		\label{eqn_l2_distortion_offset_simplified2_widthX_RVs}
	\end{equation}
	
	From (\ref{eqn_l2_distortion_offset_simplified2_widthX_RVs}), we can approximate the expected distortion. The difference between this case and $\pulseDT=1$ is that we must determine the joint PMF $\pmfFcn{\bDelayRVval{1},\bDelayRVval{2},\ldots,\bDelayRVval{\pulseLen-1},\arbitraryRVval}$ of $\pulseLen$ \emph{dependent} Binomial random variables $\{\bDelayRV{1},\bDelayRV{2},\ldots,\bDelayRV{\pulseLen-1},\arbitraryRV\}$. We derive the joint PMF using the multiplicative rule for joint probabilities, i.e.,
	\begin{align}
	\pmfFcn{\bDelayRVval{1},\bDelayRVval{2},\ldots,\bDelayRVval{\pulseLen-1},\arbitraryRVval} = &\, \pmfFcn{\bDelayRVval{2},\ldots,\bDelayRVval{\pulseLen-1},\arbitraryRVval|\bDelayRVval{1}} \pmfFcn{\bDelayRVval{1}} \nonumber \\
	= &\, \pmfFcn{\bDelayRVval{3},\ldots,\bDelayRVval{\pulseLen-1},\arbitraryRVval|\bDelayRVval{1},\bDelayRVval{2}} \nonumber \\
	& \times \pmfFcn{\bDelayRVval{2}|\bDelayRVval{1}} \pmfFcn{\bDelayRVval{1}},
	\end{align}
	and so on, where $\pmfFcn{\bDelayRVval{1}}$ is the Binomial PMF with $\targetLen-1$ trials and success probability $\probTarget$. Given knowledge of $\bDelayRVval{1}$, there are fewer trials for $\bDelayRV{2}$ (reduced to $\targetLen-1-\bDelayRVval{1}$) but the success probability increases from $(1-\probTarget)\probTarget$ to $\probTarget$. Generally, given knowledge of $\{\bDelayRVval{1},\bDelayRVval{2},\ldots,\bDelayRVval{i-1}\}$, then $\bDelayRV{i}$ is a Binomial random variable with
	$\targetLen-1-\sum_{\dPulse{}=1}^{i-1}\bDelayRVval{\dPulse{}}$ trials and success probability $\probTarget$. However, given knowledge of $\{\bDelayRVval{1},\bDelayRVval{2},\ldots,\bDelayRVval{\pulseLen-1}\}$, then the value of $\arbitraryRV$ will depend on the relative values of the filter length $\pulseLen$ and the minimum charging time $\nMin$.
	\addtocounter{equation}{1} 
	If $\pulseLen < \nMin$, then there are $\targetLen-1-\sum_{\dPulse{}=1}^{\pulseLen-1}\bDelayRVval{\dPulse{}}$ trials with success probability $(1-\probTarget)^{\nMin-\pulseLen}$. The expected distortion can then be written as in (\ref{eqn_l2_distortion_mean_widthX_L_small}) located at the top of the following page. If $\pulseLen \ge \nMin$, then $\arbitraryRV$ is precisely known since we already know the number of spikes that were separated by less than the minimum charging time, so we can write
	\begin{equation}
	\arbitraryRVval = \targetLen - 1 - \sum_{\dPulse{}=1}^{\nMin-1}\bDelayRVval{\dPulse{}},
	\end{equation}
	and
	\addtocounter{equation}{1} 
	the expected distortion simplifies to the expression shown in (\ref{eqn_l2_distortion_mean_widthX_L_big}) at the top of the following page. The evaluations of (\ref{eqn_l2_distortion_mean_widthX_L_small}) and (\ref{eqn_l2_distortion_mean_widthX_L_big}) have combinatorial complexity dependent on the length $\pulseLen$ of the filter, because we need to account for all combinations of partial overlap between the pairs of consecutive target spikes in $\targetSeq$.
	
	\begin{figure*}[!t]
		\normalsize
		\setcounter{mytempeqncnt}{\value{equation}}
		\setcounter{equation}{33}
		\begin{align}
		\distortFcnAvg{\targetSeq,\genSeq} \approx &
		\sum_{\bDelayRVval{1}=0}^{\targetLen-1} \cdots \sum_{\bDelayRVval{\pulseLen-1}=0}^{\targetLen-1-\sum_{\dPulse{}=1}^{\pulseLen-2}\bDelayRVval{\dPulse{}}}
		\sum_{\arbitraryRVval=0}^{\targetLen-1-\sum_{\dPulse{}=1}^{\pulseLen-1}\bDelayRVval{\dPulse{}}} \left[2(\targetLen-1-\arbitraryRVval)\sum_{\pulseInd=0}^{\pulseLen-1}\kernelVal{\pulseInd}^2 + 2\sum_{\dPulse{}=1}^{\pulseLen-1}\bDelayRVval{\dPulse{}}\sum_{n = \dPulse{}}^{\pulseLen-1} \kernelVal{n}\kernelVal{n-\dPulse{}}\right]^\frac{1}{2} \pmfFcn{\bDelayRVval{1},\bDelayRVval{2},\ldots,\bDelayRVval{\pulseLen-1},\arbitraryRVval} \nonumber \\
		= & \sum_{\bDelayRVval{1}=0}^{\targetLen-1} \binom{\targetLen-1}{\bDelayRVval{1}}\probTarget^{\bDelayRVval{1}}(1-\probTarget)^{\targetLen-1-\bDelayRVval{1}} \cdots \sum_{\bDelayRVval{\pulseLen-1}=0}^{\targetLen-1-\sum_{\dPulse{}=1}^{\pulseLen-2}\bDelayRVval{\dPulse{}}} \binom{\targetLen-1-\sum_{\dPulse{}=1}^{\pulseLen-2}\bDelayRVval{\dPulse{}}}{\bDelayRVval{\pulseLen-1}}\probTarget^{\bDelayRVval{\pulseLen-1}}(1-\probTarget)^{\targetLen-1-\sum_{\dPulse{}=1}^{\pulseLen-1}\bDelayRVval{\dPulse{}}} \nonumber \\
		& \times 
		\sum_{\arbitraryRVval=0}^{\targetLen-1-\sum_{\dPulse{}=1}^{\pulseLen-1}\bDelayRVval{\dPulse{}}} \left[2(\targetLen-1-\arbitraryRVval)\sum_{\pulseInd=0}^{\pulseLen-1}\kernelVal{\pulseInd}^2 + 2\sum_{\dPulse{}=1}^{\pulseLen-1}\bDelayRVval{\dPulse{}}\sum_{n = \dPulse{}}^{\pulseLen-1} \kernelVal{n}\kernelVal{n-\dPulse{}}\right]^\frac{1}{2} \nonumber \\
		& \times \binom{\targetLen-1-\sum_{\dPulse{}=1}^{\pulseLen-1}\bDelayRVval{\dPulse{}}}{\arbitraryRVval}(1-\probTarget)^{\arbitraryRVval(\nMin-\pulseLen)} \left(1-(1-\probTarget)^{\nMin-\pulseLen}\right)^{\targetLen-1-\sum_{\dPulse{}=1}^{\pulseLen-1}\bDelayRVval{\dPulse{}}-\arbitraryRVval}
		\label{eqn_l2_distortion_mean_widthX_L_small}
		\end{align}
		\setcounter{equation}{\value{mytempeqncnt}}
		\setcounter{mytempeqncnt}{\value{equation}}
		\setcounter{equation}{35}
		\begin{align}
		\distortFcnAvg{\targetSeq,\genSeq} \approx &
		\sum_{\bDelayRVval{1}=0}^{\targetLen-1} \binom{\targetLen-1}{\bDelayRVval{1}}\probTarget^{\bDelayRVval{1}}(1-\probTarget)^{\targetLen-1-\bDelayRVval{1}} \cdots \sum_{\bDelayRVval{\pulseLen-1}=0}^{\targetLen-1-\sum_{\dPulse{}=1}^{\pulseLen-2}\bDelayRVval{\dPulse{}}} \binom{\targetLen-1-\sum_{\dPulse{}=1}^{\pulseLen-2}\bDelayRVval{\dPulse{}}}{\bDelayRVval{\pulseLen-1}}\probTarget^{\bDelayRVval{\pulseLen-1}}(1-\probTarget)^{\targetLen-1-\sum_{\dPulse{}=1}^{\pulseLen-1}\bDelayRVval{\dPulse{}}} \nonumber \\
		& \times
		\left[2 \sum_{\dPulse{}=1}^{\nMin-1}\bDelayRVval{\dPulse{}}\sum_{\pulseInd=0}^{\pulseLen-1}\kernelVal{\pulseInd}^2 + 2\sum_{\dPulse{}=1}^{\pulseLen-1}\bDelayRVval{\dPulse{}}\sum_{n = \dPulse{}}^{\pulseLen-1} \kernelVal{n}\kernelVal{n-\dPulse{}}\right]^\frac{1}{2}
		\label{eqn_l2_distortion_mean_widthX_L_big}
		\end{align}
		\setcounter{equation}{\value{mytempeqncnt}}
		\hrulefill
		\vspace*{4pt}
	\end{figure*}
	
	As a simplifying special case, consider the filter length $\pulseLen = 2$, whose mean we considered in \cite{Noel2017d}. In this case, two of our assumptions are \emph{always} true: only consecutive target spikes can overlap, and if $\nMin > 1$ (i.e., if the minimum charging time is meaningful), then the density of the generated sequence $\genSeq$ will be 0. Furthermore, we will have $\pulseLen < \nMin$, unless $\nMin = 2$ (which is the smallest meaningful minimum charging time). If we write the number of target spikes that are immediately after the previous spike as $\bDelayRV{}$, then for $\pulseLen = 2$ the expected distortion expression of (\ref{eqn_l2_distortion_mean_widthX_L_small}) simplifies to
	\begin{align}
	\distortFcnAvg{\targetSeq,\genSeq} \approx & \sum_{\bDelayRVval{}=0}^{\targetLen-1} \binom{\targetLen-1}{\bDelayRVval{}}\probTarget^{\bDelayRVval{}}(1-\probTarget)^{\targetLen-1-\bDelayRVval{}} \nonumber \\
	& \times \!\!
	\sum_{\arbitraryRVval=0}^{\targetLen-1-\bDelayRVval{}}\!\!\! \left[(2\targetLen - 2 - 2\arbitraryRVval)(\kernelVal{0}^2 + \kernelVal{1}^2) + 2\kernelVal{0}\kernelVal{1}\bDelayRVval{}\right]^\frac{1}{2} \nonumber \\
	& \times \binom{\targetLen-1-\bDelayRVval{}}{\arbitraryRVval}(1-\probTarget)^{\arbitraryRVval(\nMin-2)} \nonumber \\
	& \times\left(1-(1-\probTarget)^{\nMin-2}\right)^{\targetLen-1-\bDelayRVval{}-\arbitraryRVval},
	\label{eqn_l2_distortion_mean_width2}
	\end{align}
	which uses fewer approximations than the case for general $\pulseLen$.
	
	We can use the expected distortion in (\ref{eqn_l2_distortion_mean_widthX_L_small}) or (\ref{eqn_l2_distortion_mean_widthX_L_big}) to approximate the distribution of the distortion $\distortFcnDT{\targetSeq,\genSeq}$. To use a normal approximation (as we did for the total delay in Section~\ref{sec_delay}), we only need to identify the variance $\distortVariance{\distortion}$, which we can also evaluate using (\ref{eqn_l2_distortion_mean_widthX_L_small}) or (\ref{eqn_l2_distortion_mean_widthX_L_big}). The second moment of the distortion can be found by evaluating (\ref{eqn_l2_distortion_mean_widthX_L_small}) or (\ref{eqn_l2_distortion_mean_widthX_L_big}) but removing the square root. The square of the expected distortion is found by squaring (\ref{eqn_l2_distortion_mean_widthX_L_small}) or (\ref{eqn_l2_distortion_mean_widthX_L_big}). By writing the distortion as a random variable with value $\distortRVval$, its CDF has the same form as the normal approximation of the total delay distortion in (\ref{eqn_distortion_cdf_total}).
	
	\subsection{Consideration of Infinite Kernels}
	
	A comparison between (\ref{eqn_l2_distortion_simplified}) and (\ref{eqn_l2_distortion_mean_widthX_L_small}) demonstrates the increase in complexity when there is partial overlap between filtered spikes. This approach is not suitable for infinitely-long kernels. Asymptotic analysis, where the discrete model becomes continuous as the time slot goes to 0, is of interest nevertheless. We leave such considerations for future work.
	
	\section{Numerical Results}
	\label{sec_results}
	
	In this section, we evaluate the accuracy of the distortion metrics between target and stimulated spikes that we derived in Sections~\ref{sec_delay} and~\ref{sec_filter}. We generate target sequences as a Poisson process in continuous time and as a geometric process in discrete time. To make the two time models analogous, we use discrete time slots of length $\dt=0.5\,\mathrm{ms}$ and consider a target spike probability of $\probTarget \in [10^{-3},1]$ in every slot. Since $\probTarget = \dt\rateTarget$, this corresponds to target spike generation rate range $\rateTarget \in [2,2000]\,\mathrm{s}^{-1}$. All simulations of average distortion (either delay or RMSE) are averaged over $10^4$ sequences for each $\rateTarget$ or $\probTarget$, and each simulated distortion CDF is generated by simulating $10^5$ sequences. We have confirmed that the numbers of realizations were sufficient for $95\%$ confidence intervals to be much smaller than the displayed marker size, so for clarity we do not plot confidence intervals. Unless otherwise noted, we consider a minimum charging time $\tMin=2\,\mathrm{ms}$ (i.e., $\nMin=4$ slots), which is within the range of values discussed in Section~\ref{section_model} and also comparable to a typical neuron refractory period; see \cite[Ch.~45]{Sadava2014}.
	
	\subsection{Delay-Based Distortion}
	
	To assess the delay distortion, we assume that the target sequence $\targetSeq$ has $\targetLen = 200$ spikes. When we measure the delay for a single spike (either the average or the distribution), we ignore the zero delay for the initial spike in $\targetSeq$.
	
\begin{figure}[!tb]
		\centering
		\ifOneCol
		\includegraphics[width=0.5\linewidth]{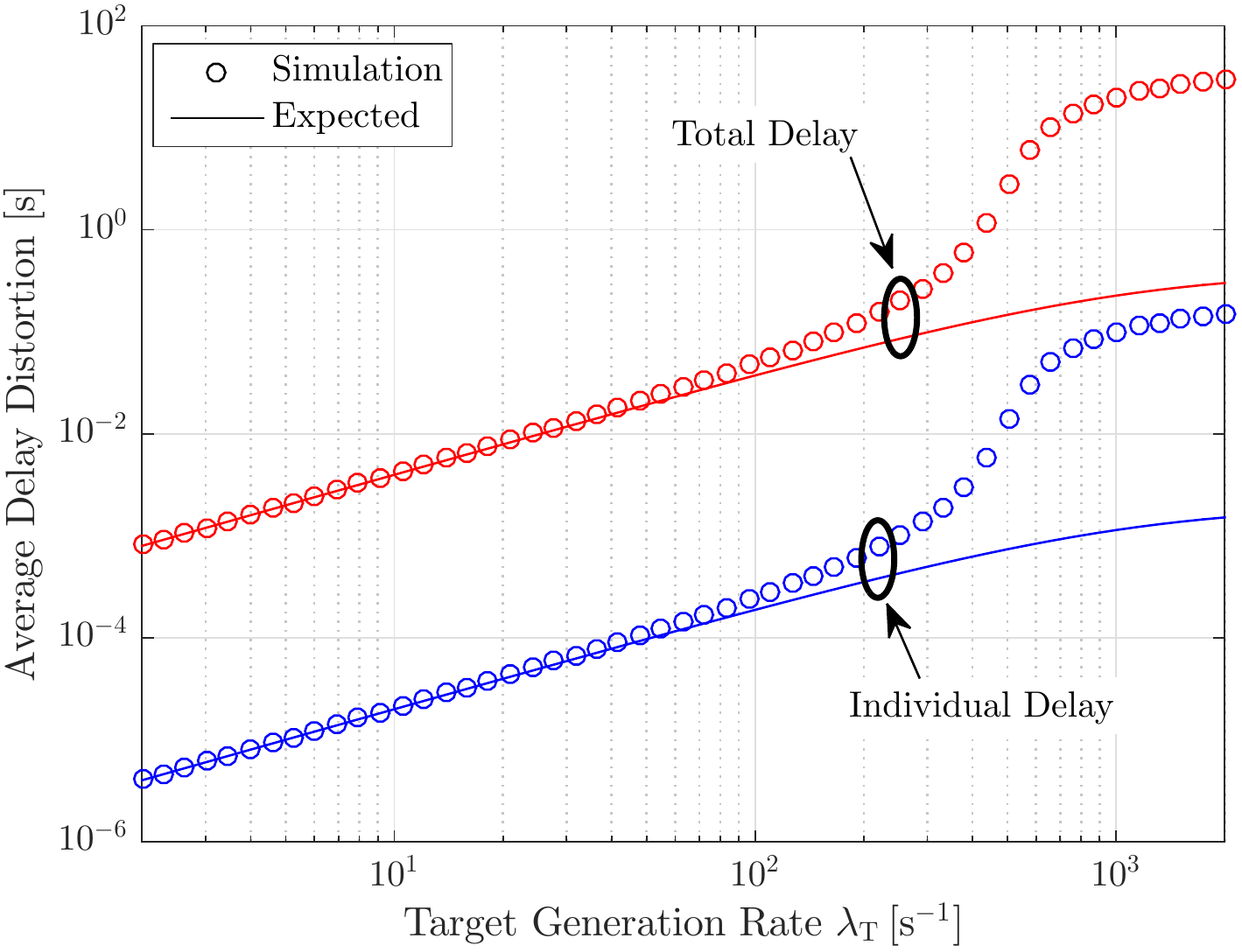}
		\else
		\includegraphics[width=\linewidth]{noel3}
		\fi
		\caption{Average delay distortion $\distortFcnAvg{\targetSeq,\genSeq}$ as a function of the target generation rate $\rateTarget$. The target sequence has a length of $\targetLen = 200$ spikes, and the minimum time that we must wait before generating another spike in $\genSeq$ is $\tMin=2\,\mathrm{ms}$. The total (cumulative) delay for a given sequence is the sum of the individual delays.}
		\label{fig_delay}
	\end{figure}

	We plot the average delay distortion in Fig.~\ref{fig_delay} on a log-log scale as a function of the target sequence generation rate $\rateTarget$. We plot the delay per pulse and the total delay per sequence of 200 pulses. The simulated distortion is found by evaluating (\ref{eqn_delay}) for each pulse. The only difference between the individual and total delay is that the average total delay is $\targetLen-1=199$ times greater than the individual delay for any $\rateTarget$.
	
	The expected analytical distortion for the delay of individual spikes and the cumulative delay plotted in Fig.~\ref{fig_delay} is calculated using the closed form expressions given in (\ref{eqn_delay_expected_evaluated}) and (\ref{eqn_delay_expected_total}), respectively. We observe that the analytical expressions are accurate for $\rateTarget < 40\,\mathrm{s}^{-1}$, i.e., when assuming a low generation rate is valid. Within this range, the average delay increases linearly with $\rateTarget$. For $\rateTarget \ge 40\,\mathrm{s}^{-1}$, the probability of more than 2 pulses within an interval of $2\tMin$ becomes non-negligible and the delay for a given spike is more likely to depend on the timing of multiple previous spikes. Thus, the expected delays, which assume that any single spike's delay is within $[0,\tMin)$, are lower bounds and become less accurate as $\rateTarget$ increases. Asymptotically, as $\rateTarget \to \infty$, it can be shown\footnote{If $\rateTarget \to \infty$, then $\targetSeq = \{0,0,\ldots,0\}$, $\genSeq = \{0,\tMin,2\tMin\ldots,(\targetLen-1)\tMin\}$, and each element in $\genSeq$ is equal to the corresponding delay.} that the average simulated delay for an individual spike (ignoring the first spike) will approach $\tMin\targetLen/2= 200\,\mathrm{ms}$. Similarly, from (\ref{eqn_delay_expected_evaluated}), the expected delay for an individual spike as $\rateTarget \to \infty$ approaches $\tMin=2\,\mathrm{ms}$.
	
\begin{figure}[!tb]
	\centering
	\ifOneCol
	\includegraphics[width=0.5\linewidth]{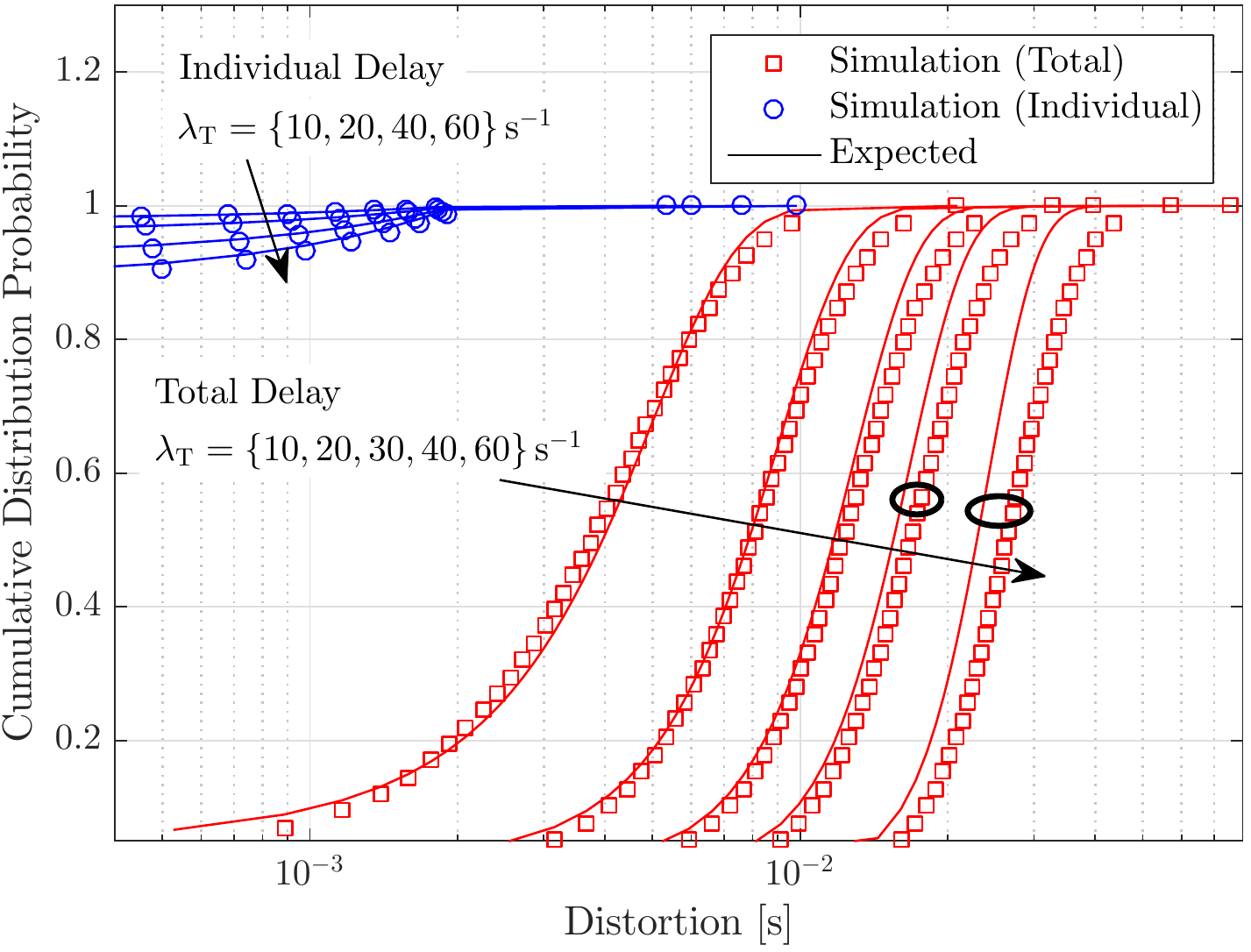}
	\else
	\includegraphics[width=\linewidth]{noel4}
	\fi
	\caption{Distribution of the delay distortion $\distortFcnAvg{\targetSeq,\genSeq}$ for different values of target generation rate $\rateTarget$. The target sequence has a length of $\targetLen = 200$ spikes, and the minimum time that we must wait before generating another spike in $\genSeq$ is $\tMin=2\,\mathrm{ms}$.}
	\label{fig_delay_distribution}
	\end{figure}

	We plot the CDF versus the delay distortion in Fig.~\ref{fig_delay_distribution} for a selection of target generation rates $\rateTarget\le 60\,\mathrm{s}^{-1}$. The expected distribution for a single spike's delay is calculated using (\ref{eqn_delay_distribution_single}), and the cumulative delay is calculated using (\ref{eqn_distortion_cdf_total}). Unlike Fig.~\ref{fig_delay}, we observe a meaningful difference in Fig.~\ref{fig_delay_distribution} between the individual and total delays. This is because the distribution of a single spike's delay is relatively narrow and there is a high probability of no delay for the generation rates considered. However, the total delays are easier to distinguish. We also observe that the normal approximation is suitable for the expected total delay, in part due to the long length $\targetLen$ of the target sequence (validating use of the central limit theorem), although its accuracy is limited as $\rateTarget$ increases.

	Overall, we have observed how the charging time places a constraint on our ability to generate spikes at arbitrary times, which also ultimately limits the amount of information that the neurons can carry. This general result is intuitive for a delay-based distortion measure, but we have presented an analytical model that enables us to make predictions about the distribution of this distortion.
	
	\subsection{Filter-Based Distortion}
	
	We now assess the filter-based distortion where we consider the RMSE metric (i.e., the $\ell^2$ norm). Due to the combinatorial complexity when calculating the expected distortion using (\ref{eqn_l2_distortion_mean_widthX_L_small}) and (\ref{eqn_l2_distortion_mean_widthX_L_big}), i.e., when $\pulseDT=2$, we reduce the target sequence length $\targetSeq$ to $\targetLen = 20$ spikes. For the filter with $\pulseDT = 1$ tap, we consider the Kronecker delta kernel so that we can apply the results from Section~\ref{sec_distortion_kernel1}. For a filter with any other number $\pulseLen$ of taps, we consider the $\pulseInd$th coefficient $\kernelVal{\pulseInd} = \pulseLen^{-\frac{1}{2}}$, so that each coefficient is weighted equally and the sum of the squares of the coefficients is equal to that of the Kronecker delta.		
	
\begin{figure}[!tb]
		\centering
		\ifOneCol
		\includegraphics[width=0.5\linewidth]{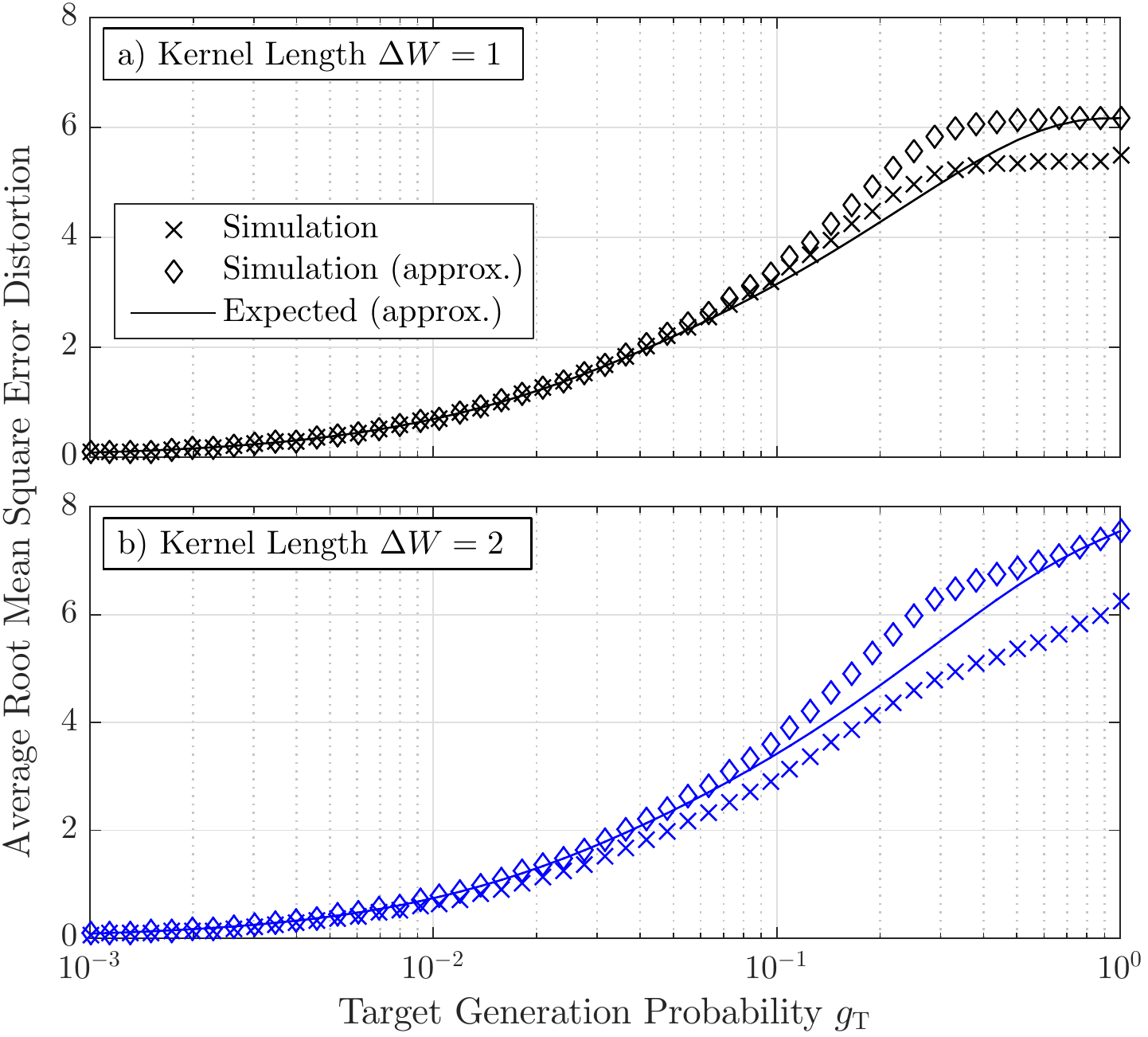}
		\else
		\includegraphics[width=\linewidth]{noel5}
		\fi
		\caption{Average RMSE $\distortFcnAvg{\targetSeq,\genSeq}$ as a function of the target firing probability $\probTarget$ in each time slot, for filter kernel length a) $\pulseDT = 1$ and b) $\pulseDT = 2$. The target sequence has a length of $\targetLen = 20$ spikes, and we are constrained by a charging time of $\nMin=4$~slots when generating spikes in $\genSeq$.}
		\label{fig_mse_average}
	\end{figure}

	In Fig.~\ref{fig_mse_average}, we measure the average RMSE as a function of the target spike generation probability $\probTarget$ for a) the filter of length $\pulseDT = 1$ and b) the filter of length $\pulseDT = 2$. We use two methods to calculate the simulated distortion. The true simulated distortion is measured using (\ref{eqn_l2_distortion_simplified}) and (\ref{eqn_l2_distortion_offset}) for the filters of length $\pulseDT = 1$ and $\pulseDT = 2$, respectively. We approximate the simulated distortion using (\ref{eqn_l2_distortion_simplified_low}) for $\pulseDT = 1$ and (\ref{eqn_l2_distortion_offset_simplified2_widthX}) for $\pulseDT = 2$, where we assume that $\probTarget$ is sufficiently low, i.e., that $\targetSeq$ is sparse. The expected analytical curves are plotted using (\ref{eqn_l2_distortion_mean}) for $\pulseDT = 1$ and (\ref{eqn_l2_distortion_mean_width2}) for $\pulseDT = 2$ and also assume that $\targetSeq$ is sparse.
	
\begin{figure}[!tb]
		\centering
		\ifOneCol
		\includegraphics[width=0.5\linewidth]{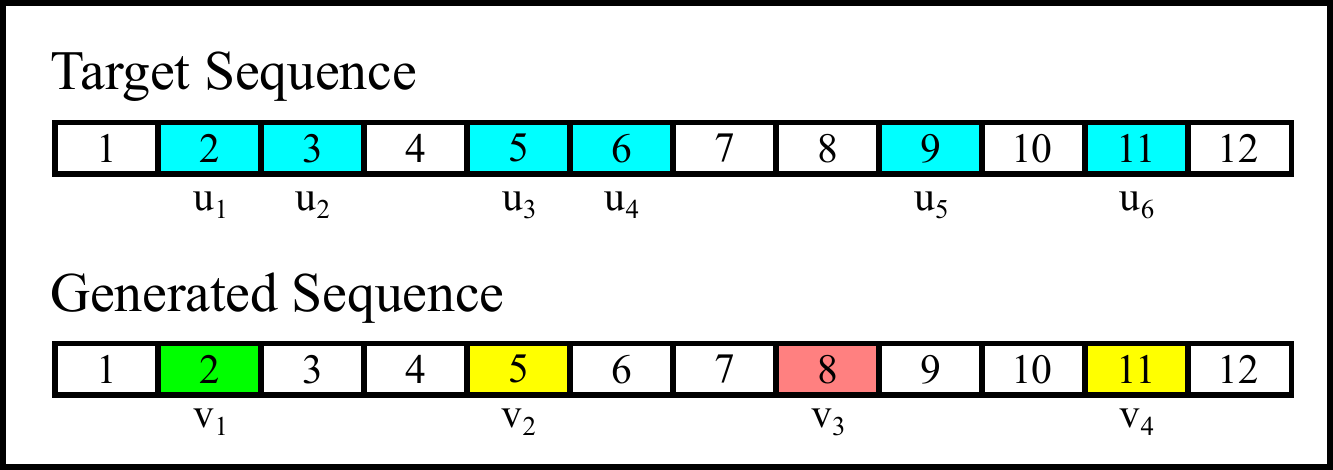}
		\else
		\includegraphics[width=\linewidth]{noel6}
		\fi
		\caption{Example of target sequence matching in discrete time, where the target sequence $\targetSeq$ has a high spike generation probability and the generated sequence $\genSeq$ is constrained by a charging time of $\nMin=3$~slots. Slots are labeled chronologically. $\targetSeq$ has 6 pulses (at start of slots shown in blue). $\genSeq$ can match the first pulse (green, in the 2nd slot), but the spikes in the $5$th and $11$th slots (yellow) match \emph{future} spikes that are not the corresponding target spikes. The spike in the $8$th slot (red) does not match any target spike. The spikes in $\genSeq$ to match $\target{5}$ and $\target{6}$ occur after the $12$th slot and are not shown.}
		\label{fig_matching_example_high_density}
	\end{figure}

	We observe in Fig.~\ref{fig_mse_average}a) that all three curves for $\pulseDT = 1$ agree well when $\probTarget < 4\,\%$, such that the timing of a given generated spike in $\genSeq$ primarily depends on the timing of only one previous spike in $\targetSeq$. For $\probTarget \ge 4\,\%$, we often have two or more spikes within an interval of $2\nMin$ slots, i.e., spikes occur sufficiently often that multiple previous spikes in $\targetSeq$ affect the timing of spikes in $\genSeq$. This generally leads to the expected distortion acting as a lower bound. However, for \emph{very} high spike generation probabilities, i.e., $\probTarget \ge 40\,\%$, the true distortion becomes \emph{lower} than that predicted by the expected curve. This is because there are so many target spikes in $\targetSeq$ that delayed spikes in $\genSeq$ are likely to occur at the same time as future spikes in $\targetSeq$, i.e., $\target{i} = \target{j} + \offset{j}$ for some $j<i$. Examples of this are shown in Fig.~\ref{fig_matching_example_high_density}. Such occurrences are not accounted for in the derivations of the approximations, where spikes in $\genSeq$ must align with the \emph{corresponding} spikes in $\targetSeq$, but from (\ref{eqn_l2_distortion_simplified}) these asynchronous overlaps lead to a smaller true distortion, which saturates as $\probTarget \to 1$. Thus, the expected distortion is a lower bound in the ``low density'' regime but an upper bound in the ``high density'' regime. We also note that the approximate simulated distortion converges to the expected distortion as $\probTarget \to 1$. This is because \emph{every} spike generated after the initial one has \emph{no overlap} with its corresponding target spike, so both distortion measures are maximized to the same value, i.e., $\distortFcnAvg{\targetSeq,\genSeq} = \sqrt{2\targetLen-2} = \sqrt{38}$.
	
	For $\pulseDT = 2$, Fig.~\ref{fig_mse_average}b) also shows that the expected distortion is a lower bound on the approximate simulated distortion and all distortions saturate as $\probTarget \to 1$. The lower bound is accurate for low $\probTarget$ (here when $\probTarget < 5\,\%$) and then converges when $\probTarget \to 1$. However, unlike the $\pulseDT = 1$ case, we see that the expected distortion for $\pulseDT = 2$ is an upper bound on the true distortion for \emph{all} $\probTarget$. This is a side effect of the longer filter; non-zero overlap occurs between a target spike in $\targetSeq$ and the corresponding spike in $\genSeq$ when the latter is generated with a one-slot delay. Such ``imperfect'' overlap reduces the measure of distortion calculated using (\ref{eqn_l2_distortion_offset}) but is not accounted for in the approximate or expected distortion.
	
\begin{figure}[!tb]
		\centering
		\ifOneCol
		\includegraphics[width=0.5\linewidth]{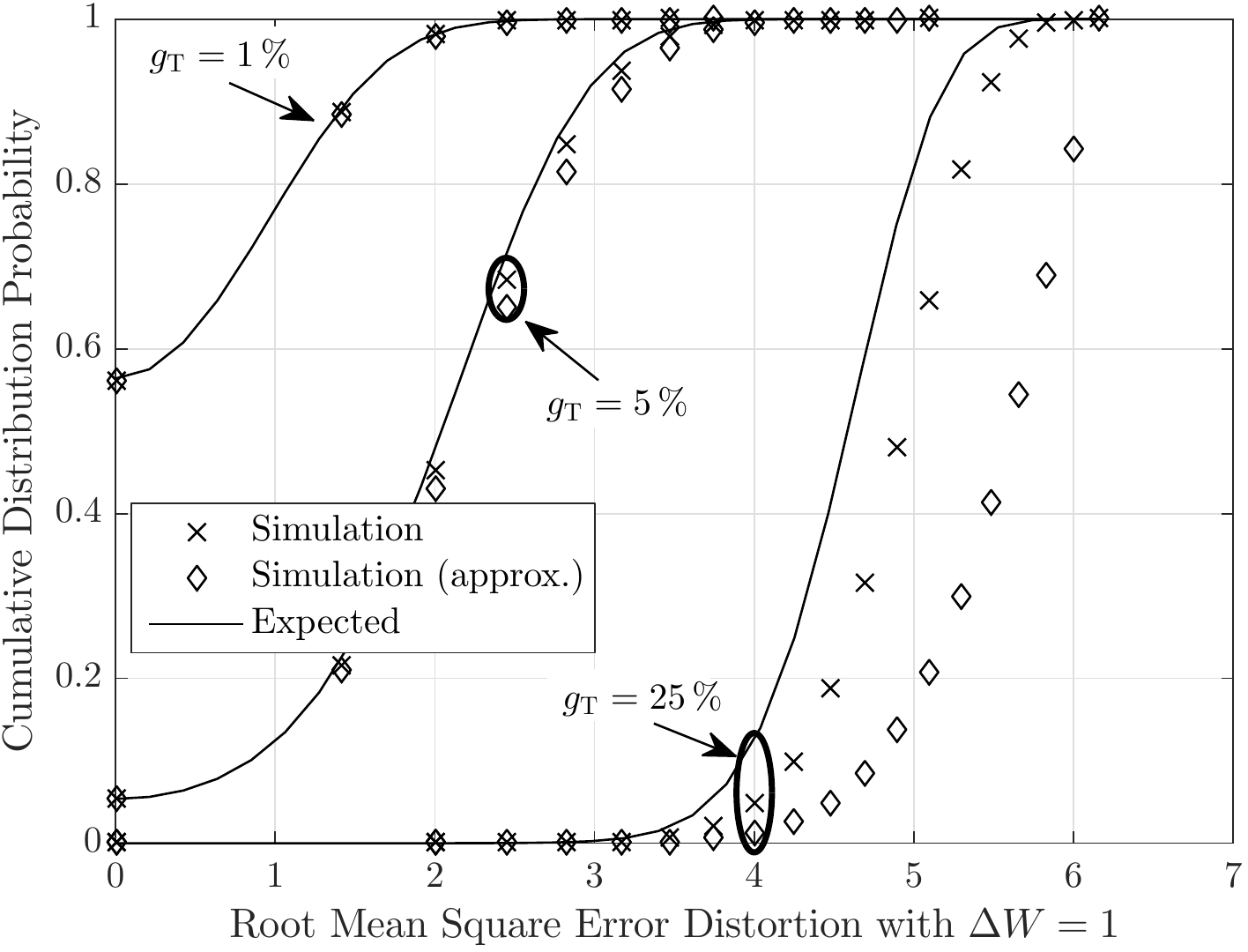}
		\else
		\includegraphics[width=\linewidth]{noel7}
		\fi
		\caption{Distribution of RMSE $\distortFcnDT{\targetSeq,\genSeq}$ when the metric kernel has length $\pulseDT=1$. Target firing probabilities of $\probTarget = \{1,5,25\}\%$ are considered. The target sequence has a length of $\targetLen = 20$ spikes, and we are constrained by a charging time of $\nMin=4$~slots when generating spikes in $\genSeq$.}
		\label{fig_distortion_mse_length1_distribution}
	\end{figure}

	Now that we have assessed the average RMSE distortion, we consider its distribution. We plot the distribution of the RMSE for the filter of length $\pulseDT = 1$ in Fig.~\ref{fig_distortion_mse_length1_distribution}. We consider the CDF for three values of target spike generation probability, i.e., $\probTarget = \{1,5,25\}\%$. The simulated distortion distributions are generated from the same equations used to observe the average distortion in Fig.~\ref{fig_mse_average}a), i.e., (\ref{eqn_l2_distortion_simplified}) and (\ref{eqn_l2_distortion_simplified_low}) are used to measure the true and approximate distortions, respectively. The expected CDF is calculated using (\ref{eqn_l2_distortion_distribution}). In Fig.~\ref{fig_distortion_mse_length1_distribution}, we observe good agreement between the true RMSE distribution and that observed and calculated assuming low $\probTarget$ when $\probTarget=1\,\%$. Slight deviations from the true distribution are observed for $\probTarget=5\,\%$, and larger deviations are observed for $\probTarget=25\,\%$. Overall, the accuracy of the approximations is consistent with that observed for the average distortion in Fig.~\ref{fig_mse_average}a). If we also included the distortion CDFs for $\probTarget=100\,\%$, then we would observe that they are step functions that transition at the corresponding ``average'' distortions in Fig.~\ref{fig_mse_average}a). This is because the target sequence (and correspondingly the generated sequence) becomes \emph{deterministic} in this case, with a spike in each of the first $\targetLen$ time slots.
	
\begin{figure}[!tb]
	\centering
	\ifOneCol
	\includegraphics[width=0.5\linewidth]{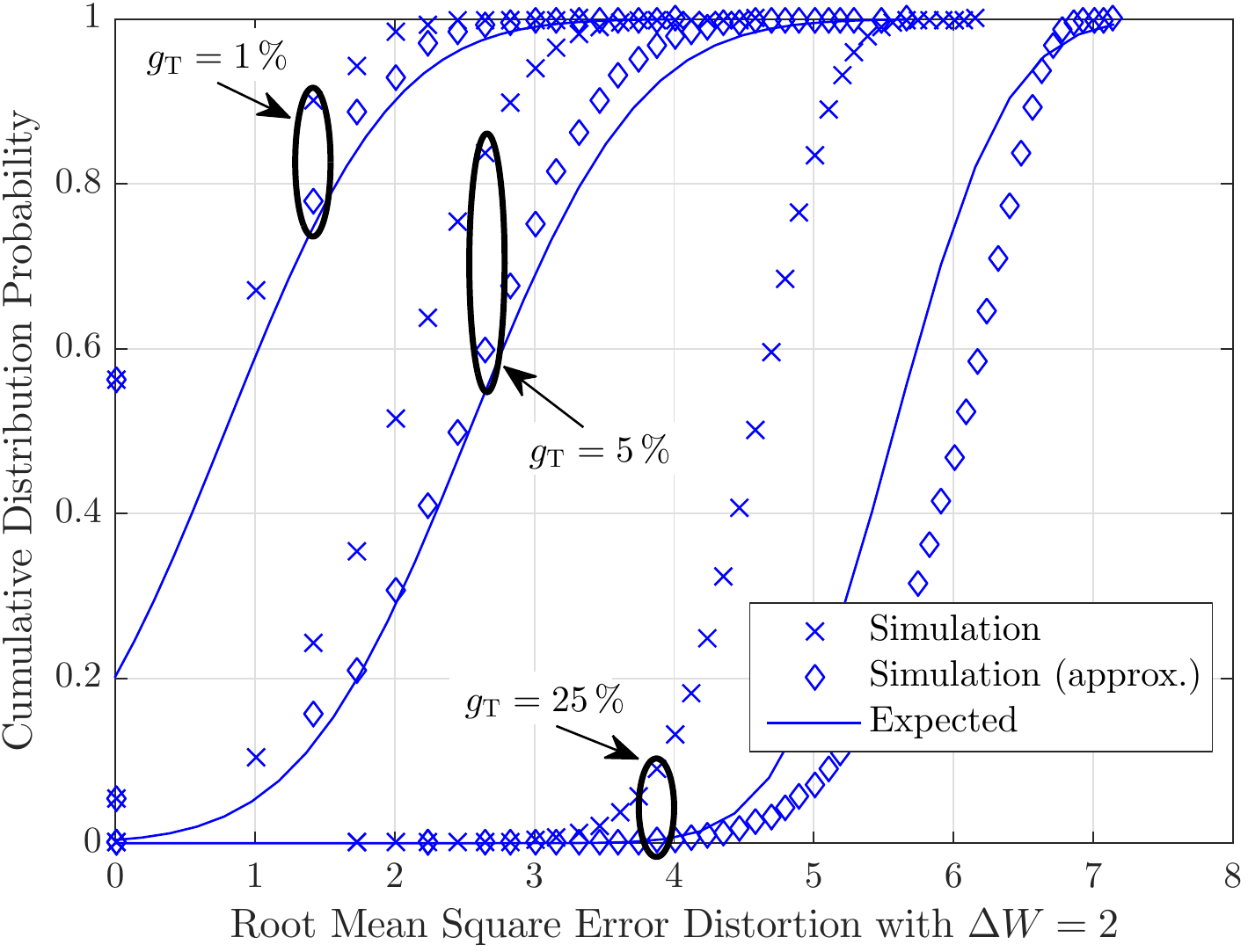}
	\else
	\includegraphics[width=\linewidth]{noel8}
	\fi
	\caption{Distribution of RMSE $\distortFcnDT{\targetSeq,\genSeq}$ when the metric kernel has length $\pulseDT=2$. Target firing probabilities of $\probTarget = \{1,5,25\}\%$ are considered. The target sequence has a length of $\targetLen = 20$ spikes, and we are constrained by a charging time of $\nMin=4$~slots when generating spikes in $\genSeq$.}
	\label{fig_distortion_mse_length2_distribution}
	\end{figure}

	Fig.~\ref{fig_distortion_mse_length2_distribution} plots the distribution of the RMSE for the filter of length $\pulseDT = 2$ with target generation probabilities $\probTarget = \{1,5,25\}\%$. The simulated distortion distributions are generated from the same equations used to observe the average distortion in Fig.~\ref{fig_mse_average}b), i.e., (\ref{eqn_l2_distortion_offset}) and (\ref{eqn_l2_distortion_offset_simplified2_widthX}) are used to measure the true and approximate distortions, respectively. The expected CDF is calculated using (\ref{eqn_distortion_cdf_total}), where the mean and standard deviation are calculated using (\ref{eqn_l2_distortion_mean_width2}). The deviations between the simulated and expected distributions are larger than they are for the case of $\pulseDT = 1$, as a result of less accuracy in the average RMSE and also due to the normal approximation. Analogously to the discussion for Fig.~\ref{fig_distortion_mse_length1_distribution}, distortion CDFs for $\probTarget=100\,\%$ would appear as step functions corresponding to the ``average'' values in Fig.~\ref{fig_mse_average}b).

\begin{figure}[!tb]
	\centering
	\ifOneCol
	\includegraphics[width=0.5\linewidth]{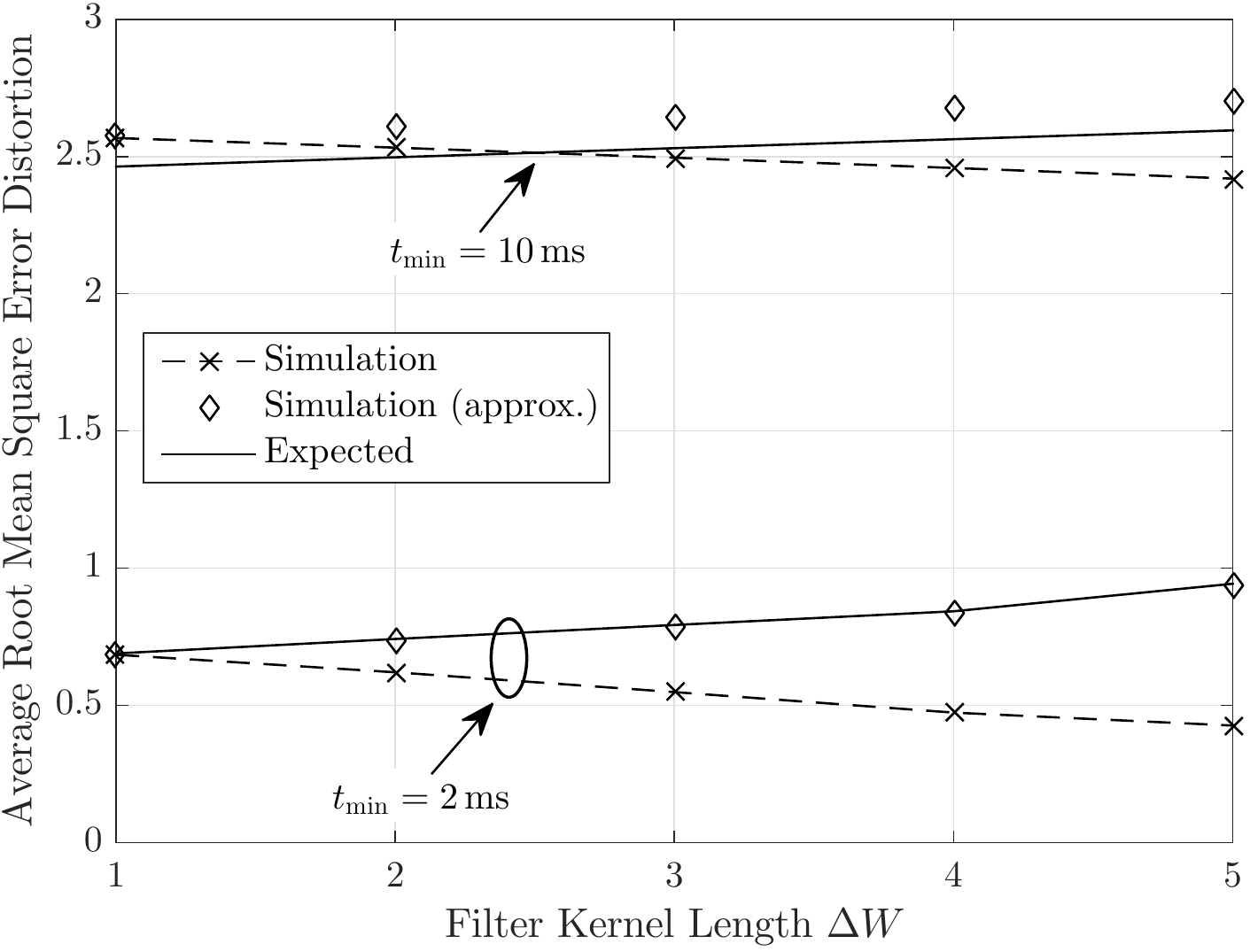}
	\else
	\includegraphics[width=\linewidth]{noel9}
	\fi
	\caption{Average RMSE $\distortFcnAvg{\targetSeq,\genSeq}$ as a function of the filter length $\pulseDT$. A target firing probability of $\probTarget = 1\,\%$ is considered. The target sequence has a length of $\targetLen = 20$ spikes, and we vary the charging time $\tMin$. The minimum wait times $\tMin = \{2,10\}\,\mathrm{ms}$ correspond to $\nMin = \{4,20\}$~slots. The actual simulated $\distortFcnAvg{\targetSeq,\genSeq}$ decreases with increasing $\pulseDT$, whereas the approximation of $\distortFcnAvg{\targetSeq,\genSeq}$ (and its expected value) increases with increasing $\pulseDT$.}
	\label{fig_mse_vs_length}
	\end{figure}

	We consider longer filter lengths in Fig.~\ref{fig_mse_vs_length}, where we plot the average RMSE as a function of the filter length $\pulseDT$ for different minimum charging times $\tMin = \{2,10\}\,\mathrm{ms}$, which correspond here in discrete time to $\nMin = \{4,20\}$~slots. The target firing probability is $\probTarget = 1\,\%$, which is sparse for relatively low values of $\nMin$. For $\pulseDT>2$, the expected analytical points are calculated using (\ref{eqn_l2_distortion_mean_widthX_L_big}) or (\ref{eqn_l2_distortion_mean_widthX_L_small}) (depending on whether $\pulseDT = \pulseLen \ge \nMin$, respectively). The expected points are calculated using (\ref{eqn_l2_distortion_mean}) and (\ref{eqn_l2_distortion_mean_width2}) for $\pulseDT = 1$ and $\pulseDT = 2$, respectively. The actual simulated distortion is calculated using (\ref{eqn_l2_distortion_simplified}) and (\ref{eqn_l2_distortion_offset}), and the approximate simulated distortion is calculated using (\ref{eqn_l2_distortion_simplified_low}) and (\ref{eqn_l2_distortion_offset_simplified2_widthX}), for $\pulseDT = 1$ and $\pulseDT \ge 2$, respectively. For $\tMin = 2\,\mathrm{ms}$, the expected distortion is in good agreement with the approximate simulated distortion, which verifies both (\ref{eqn_l2_distortion_mean_widthX_L_big}) and (\ref{eqn_l2_distortion_mean_widthX_L_small}). Interestingly, we observe that the actual simulated distortion decreases with increasing $\pulseDT$, whereas the expected and approximate simulated distortion increases with increasing $\pulseDT$. This is because a longer filter leads to a greater chance of filtered sequences partially overlapping, but also imposes a greater separation of pulses for the sparsity assumption to be valid.

\begin{figure}[!tb]
		\centering
		\ifOneCol
		\includegraphics[width=0.5\linewidth]{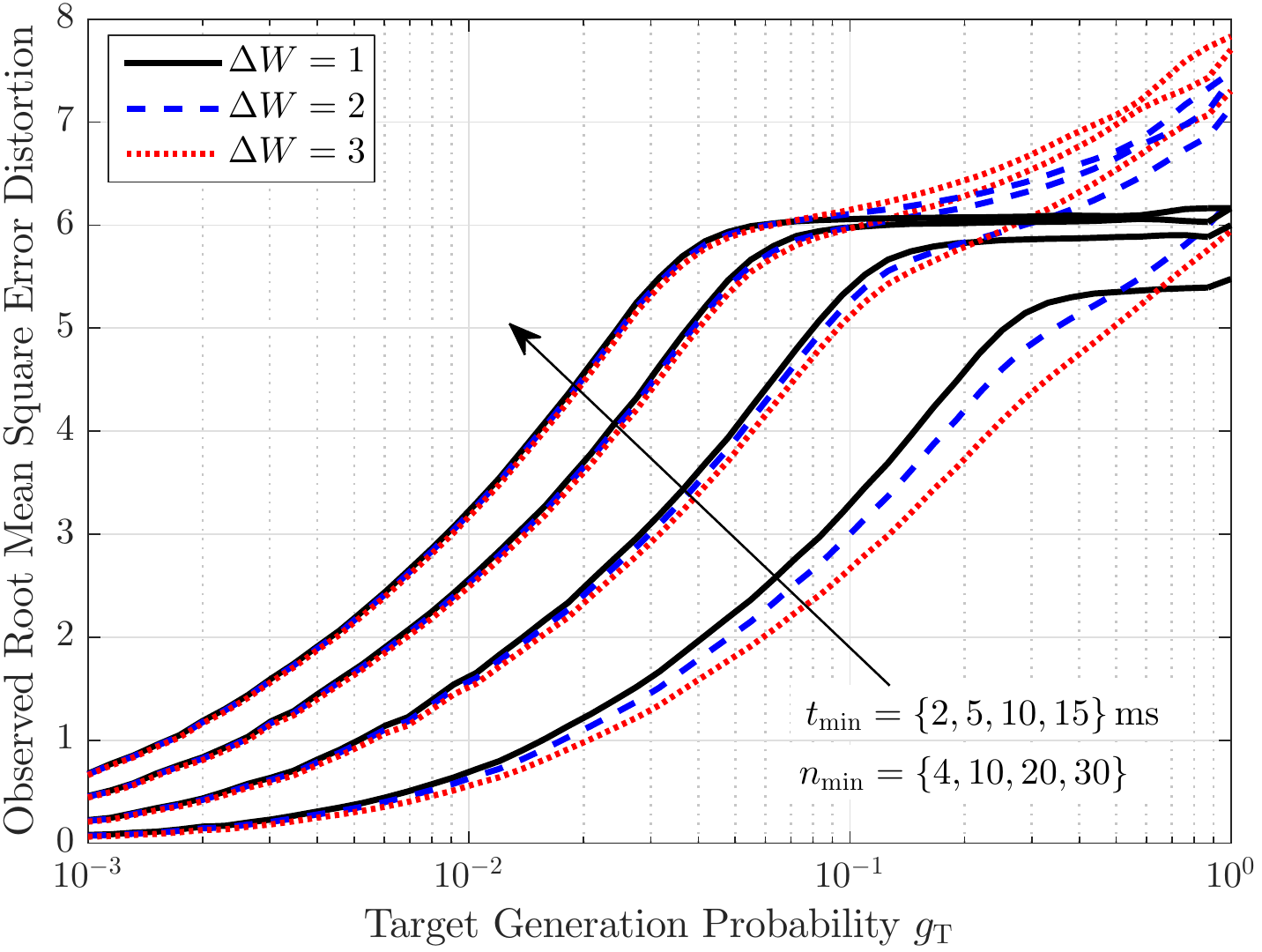}
		\else
		\includegraphics[width=\linewidth]{noel10}
		\fi
		\caption{Average observed (i.e., simulated) RMSE $\distortFcnAvg{\targetSeq,\genSeq}$ as a function of the target firing probability $\probTarget$ in each time slot. We vary the number $\nMin$ of slots that it takes to charge the neuron, and consider metric kernels with lengths $\pulseDT\in\{1,2,3\}$ slots. The minimum wait times $\tMin = \{2,5,10,15\}\,\mathrm{ms}$ correspond to $\nMin = \{4,10,20,30\}$~slots.}
		\label{fig_distortion_diff_tmin}
	\end{figure}

	Finally, in Fig.~\ref{fig_distortion_diff_tmin}, we measure the RMSE as a function of the target spike generation probability $\probTarget$ for various minimum charging times and filter lengths. Specifically, we consider charging times $\tMin = \{2,5,10,15\}\,\mathrm{ms}$, which correspond to $\nMin = \{2,4,10,20\}$~slots, and test filter lengths $\pulseDT\in\{1,2,3\}$. To facilitate the comparison, we only plot the true RMSE for each filter (i.e., using (\ref{eqn_l2_distortion_simplified}) for $\pulseDT=1$ and  (\ref{eqn_l2_distortion_offset}) for $\pulseDT\in\{2,3\}$).
	
	In Fig.~\ref{fig_distortion_diff_tmin}, we see that for lower target firing probabilities, i.e., $\probTarget < 10\%$, the RMSEs of the $\pulseDT\in\{2,3\}$ filters approach that of the length 1 filter as $\nMin$ increases. This should not be surprising, since the relative length of the filter decreases as $\nMin$ increases, and we defined the sum of the square of the coefficients to be the same for each filter. Over this range of $\probTarget$, the RMSE also decreases with increasing filter length, as we observed in Fig.~\ref{fig_mse_vs_length}. The reason for this is the same reason why the expected distortion was an upper bound on the observed distortion in Fig.~\ref{fig_mse_average}; partial overlap between spikes in $\targetSeq$ and $\genSeq$ reduces the measure of distortion but partial overlap cannot occur when the filter length is only 1 slot. Interestingly, for each $\nMin$ there is a $\probTarget$ beyond which the RMSEs with filter lengths $\pulseDT\in\{2,3\}$ become larger than that with filter length 1. We can understand this from (\ref{eqn_l2_distortion_offset}); the partial overlap between spikes in $\targetSeq$ and $\genSeq$, which reduces distortion, becomes less likely with increasing $\nMin$, whereas the density of $\targetSeq$, which increases distortion, increases with $\probTarget$. Thus, the RMSEs for filter lengths $\pulseDT\in\{2,3\}$ continue to increase with increasing $\probTarget$, whereas the RMSE for filter length 1 saturates.
	
	The filter-based distortion results	give us insight into the importance of the filter length. For a sufficiently low target spike generation rate, a longer filter places less emphasis on the precise timing of the spikes. However, this observation is not reflected in the simplified distortion model and our analytical results (see Fig.~\ref{fig_mse_vs_length}). Further work is required to effectively account for these dynamics. Also, as with the delay-based distortion, we observe the impact of the charging time as a constraint on matching target sequences.
	
	\section{Conclusions}
	\label{sec_conclusions}
	
	In this paper, we used a simple optogenetic model for externally stimulating a neuron and generating spike trains. Given a constraint on the neuron's charging time, we measured the distortion in a spike train from a target train. We measured the distortion as either the delay in generating a spike or as the filtered train's RMSE from the filtered target sequence. We derived both the mean and the distribution of the distortion under the assumption that the spike generation rate in the target sequence is sufficiently low.
	
	This is a preliminary work to understand the information carried in an externally-generated sequence of neuron pulses.	We seek to model how well pulse sequences with sufficiently small deviations can carry the same information. For example, with experimental neuron firing data (such as that in \cite{Narayan2006}), and an appropriate distortion measure and distortion threshold, we might become able predict the likelihood that 2 given neuron pulse sequences were in response to the same stimulus.
	
	There are also a number of opportunities to expand our analysis. For example, we may adopt more biologically plausible membrane potential models, including those reviewed in \cite{Izhikevich2004}, we may use more realistic optogenetic models, such as those in \cite{Nikolic2009}, and we may consider alternative metrics to compare sequences, such as those reviewed in \cite{Dauwels2008}. Our analysis only considered the stimulation of the sensory neuron, but our approach for comparing sequences could also be applied to study the propagation of spikes either along a neuron or to other neurons.

	
\bibliography{NoelEckfordDistortion}
	
\end{document}